\documentclass[twocolumn]{aastex63}
\usepackage{amsmath}
\usepackage[title]{appendix}
\usepackage{footnote}
\PassOptionsToPackage{unicode}{hyperref}
\usepackage{graphicx}
\usepackage[flushleft]{threeparttable}

\accepted{December 7, 2020}
\shorttitle{Physical and Chemical Structure of L1527}
\shortauthors{Flores-Rivera et al. 2020}
\graphicspath{{./}{figures/}}
\begin{document}

\title{Physical and Chemical Structure of the Disk and Envelope of the Class 0/I protostar L1527}

\correspondingauthor{Lizxandra Flores-Rivera \& Susan Terebey}
\email{lflor108@calstatela.edu, sterebe@calstatela.edu}

\author[0000-0001-8292-1943]{Lizxandra Flores-Rivera}
\affil{Department of Physics \& Astronomy, California State University at Los Angeles, Los Angeles, CA 90031, USA}
\affiliation{Jet Propulsion Laboratory, California Institute of Technology, Pasadena, CA 91109, USA}
\affiliation{Max-Planck Institute for Astronomy, K$\ddot{o}$nigstuhl 17, 69117 Heidelberg, Germany}

\author{Susan Terebey}
\affiliation{Department of Physics \& Astronomy, California State University at Los Angeles, Los Angeles, CA 90031, USA}
\affiliation{Jet Propulsion Laboratory, California Institute of Technology, Pasadena, CA 91109, USA}

\author{Karen Willacy}
\affiliation{Jet Propulsion Laboratory, California Institute of Technology, Pasadena, CA 91109, USA}

\author{Andrea Isella} \affiliation{Department of Physics \& Astronomy,
Rice University, 6100 Main Street
Houston, TX, 77005} 

\author{Neal Turner}
\affiliation{Jet Propulsion Laboratory, California Institute of Technology, Pasadena, CA 91109, USA}

\author{Mario Flock}
\affiliation{Jet Propulsion Laboratory, California Institute of Technology, Pasadena, CA 91109, USA}
\affiliation{Max-Planck Institute for Astronomy, K$\ddot{o}$nigstuhl 17, 69117 Heidelberg, Germany}

\begin{abstract}
Sub-millimeter spectral line and continuum emission from the protoplanetary disks and envelopes of protostars are powerful probes of their structure, chemistry, and dynamics.  Here we present a benchmark study of our modeling code, RadChemT, that for the first time uses a chemical model to reproduce ALMA C$^{18}$O (2-1) and CARMA $^{12}$CO (1-0) and N$_{2}$H$^{+}$ (1-0) observations of L1527, that allow us to distinguish the disk, the infalling envelope and outflow of this Class 0/I protostar.  RadChemT combines dynamics, radiative transfer, gas chemistry and gas-grain reactions to generate models which can be directly compared with observations for individual protostars. Rather than individually fit abundances to a large number of free parameters, we aim to best match the spectral line maps by (i) adopting a physical model based on density structure and luminosity derived primarily from previous work that fit SED and 2D imaging data, updating it to include a narrow jet detected in CARMA and ALMA data near ($\leq 75$au) the protostar, and then (ii) computing the resulting astrochemical abundances for 292 chemical species.

Our model reproduces the C$^{18}$O and N$_{2}$H$^{+}$ line strengths within a factor of 3.0; this is encouraging considering the pronounced abundance variation (factor $> 10^3$) between the outflow shell and CO snowline region near the midplane. Further, our modeling confirms suggestions regarding the anti-correlation between N$_{2}$H$^{+}$ and the CO snowline between 400 au to 2,000 au from the central star. Our modeling tools represent a new and powerful capability with which to exploit the richness of spectral line imaging provided by modern submillimeter interferometers.

\end{abstract}


\section{Introduction} 
\label{sec:intro}

In star formation theory, a low-mass protostar in its earliest stage of order of $10^{4}$ years, is considered a Class 0 newborn star embedded in a dense core typically containing a mass density of $\sim$10$^{-19}$ g~cm$^{-3}$ \citep{Bergin_2007}. During the rotational-collapse process at around 10$^{5}$ years, the cloud conserves angular momentum where the low-angular-momentum parts form a protostar, while the higher-angular-momentum parts settle into a disk in orbit around the protostar \citep{Terebey_1984}; where the gas and dust spiral inward through the disk and accrete onto the protostar.  

Magnetic fields embedded in the initial cloud core can resist the collapse, and remove angular momentum by carrying the rotational energy of the core away into the surrounding molecular cloud. However, if the cloud is weakly-ionized, its coupling to magnetic fields is slight, and magnetic fields can be ignored in the collapse. Several recent reviews suggest that magnetic fields may play a relatively small role during cloud collapse \citep{Hull_2019,Krumholz_2019}. When magnetic fields are negligible, the collapse conserves angular momentum along streamlines.  The density structure in an initially-quiescent, isothermal, collapsing core can then be described
by the Ulrich \citep{Ulrich_1976,Cassen_1981} envelope model, hereafter UCM collapse model, or the inside-out Terebey, Shu, Cassen, hereafter TSC collapse model \citep{Terebey_1984}. These two models differ from each other in terms of gas pressure support and infall velocity, particularly outside the collapse radius ($\sim 4000$~au, see \S \ref{sec:envelope_prescription}).  The whole picture of the protostellar system can be inferred from dust continuum and spectral line emission, where the age, luminosity, and mass of young protostar can be determined \citep{Kenyon_1995,Evans_2009, Dunham_2012, J_rgensen_2013, Frimann_2016}.

Predicting chemical abundances in envelopes and disks of protostars provides a way to trace the chemical inheritance of water and organics in planet formation; however, carrying out a realistic time-dependent chemical model using 2D physical properties is technically challenging. An early chemical study by \citet{ceccarelli_1996} assumed 1D (radial) physical structure for the envelope, and by determining temperature via radiative transfer, concluded that water would be abundant at temperatures above $100$K in protostellar envelopes. Other studies, also using 1D physical models with radiative transfer, further demonstrated the importance of temperature, from showing extreme CO depletion in cold outer envelopes, to the presence of simple organic molecules in corinos, namely regions of warm gas near the embedded protostar  \citep{jorgensen_2002, jorgensen_2005, ceccarelli_2006, Yang_2020}. Time dependence in the chemistry has also been used to predict chemical signatures that result from episodic accretion bursts in protostars \citep{jorgensen_2015,Visser_2015,rab_2017}.

Subsequent studies have extended physical modeling from 1D to 2D \citep{Visser_2009, Visser_2011, Drozdovskaya_2014, Drozdovskaya_2016} and focus on the time history of a parcel of gas, in order to predict the abundance of species such as CO, water, and methanol within disks. Their conclusions on chemistry, while informative, are not compared with observations of protostars, and moreover, make specific assumptions about luminosity evolution that are difficult to test.

However, for comparision with observation, the previous studies have limitations, and in particular, they do not model warm gas that is associated with outflows. Many molecules show strong emission from gas in outflows. For example, \citet{Kristensen_2017} present data for a protostar sample observed with Herschel; the velocity data for water and high-J CO lines show that these spectral lines are clearly associated with the outflow. In order to more reliably model molecules in the different spatial regions, a fully 2D description that includes outflow dynamics is necessary. In this contribution, our goal is to develop a fully 2D code that self-consistently predicts abundances for many molecular species in protostars, and moreover that can be compared with high spatial resolution spectral line data from ALMA. We try to be parsimonious in the model, to generate realistic results that include a simple outflow description, and that also minimize the inclusion of luminosity evolution, in order to better understand the importance of these effects. In this paper, we do not intend to find the best fitting parameters for the protostar L1527 but rather use what has been previously determined by \citet{tobin_2012} as initial parameters to explore the chemistry of the envelope in L1527 using RadChemT, which combines a physical and chemical model (see \S\ref{sec:parameters} and \S\ref{sec:luminosity} for more details). 

We use the protostar L1527 to compare and validate our chemical model calculations produced by RadChemT (see \S\ref{sec:overview} for more details of the utility of this package). L1527 is a typical dense core system classified as a class 0/I protostar with an edge-on disk orientation located 140 parsecs away in Taurus molecular cloud \citep{Beichman_1986, Torres_2007, Tobin_2008, Kristensen_2012}. The edge-on orientation is favorable for studying the velocity structure of the envelope and outflow that is important to later constrain the mass of the protostar itself. Previous effort by \citet{Tobin_2008, Tobin_2010, tobin_2012} provided extensive fitting, through modeling and imaging, for the physical properties of the system, although the non-unique nature of SED fitting \citep{Robitaille_2006} means that there is still a wide range in free parameters to play with. More recently, \citet{Aso_2017} constrained the radius and mass of L1527 using high resolution data from Atacama Large Millimiter/submillimiter Array (ALMA).  

Using RadChemT, we attempt to simultaneously match the observed C$^{18}$O(2-1), $^{12}$CO(1-0), and N$_{2}$H$^{+}$(1-0) line strength and PV diagrams, together with continuum data at infrared through millimeter wavelengths. These molecules and transitions are of interest since they are widely detectable by sub-millimeter interferometers. $^{12}$CO is an abundant species having strong transitions, so it traces even the low-density gas in the outflow. C$^{18}$O probes higher gas density ($\geq$ 10$^{7}$ cm$^{-3}$) regions that are affected by photodissociation processes caused by UltraViolet (UV) photons close to the disk surface and is a very good tracer of the gas properties, structure and kinematics \citep{Visser_2009}. The rotational line emission of C$^{18}$O has a much lower optical depth than $^{12}$CO, thus probing in more detail the inner parts of the envelope \citep{Henning_2013, rab_2017}. Observations for some class I systems  \citep{Takakuwa_2012, Takakuwa_2013, Takakuwa_2017, Yen_2013, Murillo_2015, Harsono_2013} have shown that gaseous disks can be Keplerian, but understanding their formation relies highly on what the spectral line emission is telling us. Moreover, non-Keplerian dynamics occurs in the cold outer disk boundary where the gas motion is free-falling onto the disk, shocks are expected, and there is the potential for a different structural gas flow \citep{Sakai_2014}. N$_{2}$H$^{+}$ is one of the last species depleted in prestellar cores, and traces the coldest, densest material ($\sim$10$^{7}$ cm$^{-3}$).

We present the first chemical calculations for L1527 using RadChemT. We provide a useful description of the physical parameters and the chemical model used to support the observations by \textit{Spitzer, Herschel}, CARMA and ALMA. As for the physical model description, we compare the UCM and the TSC collapse models and choose a density distribution to decide which one is more compelling to describe what is happening at the line center of the spectral data. We use the \textsc{Hochunk3d} code (\citet{Whitney_2013}) to determine the temperature using continuum Monte Carlo Radiative Transfer (MCRT). Then, we evolve a chemical network at each cell to find the molecular abundances, evaluating the regions dominated by the UV field and, finally, construct synthetic spectral lines to compare against the interferometric dataset. In \S\ref{sec:physicalmodel} we describe in detail the set up of the physical model to simulate L1527 observations. In \S\ref{sec:chemicalmodel} we present the essentials to model the chemistry based on the temperature and density gathered from the physical model. In \S\ref{sec:modelvisualization} we present the tools to produce images of L1527. In \S\ref{sec:observations}, we present the observational parameters of ALMA and CARMA. In \S\ref{sec:results} we present the results and discussions of the physical structure and chemical evolution using the tracers already mentioned earlier for L1527, and outlining the new CARMA data together with ALMA data for validation of the models. Finally, in  \S\ref{sec:conclusions} we present a summary of our findings.

\section{Methods} \label{sec:methods}

\subsection{RadChemT overview}
\label{sec:overview}

To make the modeling effort tractable, the software package we have assembled, RadChemT,  breaks the calculation into three parts with several key simplifying assumptions. Step 1 carries out MCRT using \textsc{Hochunk3d}, as described in \S\ref{sec:physicalmodel}, to calculate dust temperature based on the current-time luminosity $L_{\rm int}$ and 2D input physical model. Namely, for a given snapshot in time, both density $\rho(r,\theta,\phi)$ and velocity $ \bf{v}(r,\theta,\phi)$ are specified for the main components during the dynamical collapse of the class~0/I system: outflow, rotationally flattened envelope, and disk. The physical model is tuned to L1527, based on fitting modeling and imaging data from \citet{Tobin_2008, Tobin_2010, tobin_2012}.

Step 2 performs a full time-dependent chemical abundance calculation in a point model, i.e. locally for each cell as described in \S\ref{sec:chemicalmodel}, using the density and temperature from Step 1. As a simplification, this first version of RadChemT does not track material along a gas streamline as a function of time. Instead, the abundance calculations at each spatial grid point are time-dependent but assume fixed density and temperature. The assumption is that $\it current$ local conditions $(\rho, T)$ sufficiently describe earlier times. The assumption is most valid in the outer envelope, where density and temperature change the least for a moving gas parcel. There is an initial (brief) transient behavior in the chemistry, from starting with elements in atomic form, until molecular cloud abundances are achieved. After this, the chemical age should be the same as the protostar age. The age of an individual protostar can be estimated using $t_{age}=M_*/\dot{M}_{\rm env}$, but the age of an individual object is not well known during this early stage. We therefore compare chemical abundance and morphology at two time stamps, $10^4$ and $10^5$ years, that roughly span the range of relevant ages for the Class 0/I phase. Predicted abundance that are relatively constant over that time span give confidence that conclusions based on those abundances will be robust. 

Step 3 visualizes the protostar in a Position-Velocity (PV) cube, making use of the Doppler effect in spectral line emission to sense the 3D velocity structure. This investigation uses \textsc{RADMC-3D}, as described in \S\ref{sec:modelvisualization}, to perform spectral line radiative transfer assuming LTE. The main inputs are the density, temperature, and velocity grids from Step 1, combined with an abundance grid from Step 2 for a selected molecule at $t = 10^5$ years. From the inputs \textsc{RADMC-3D} creates a synthetic model PV cube in FITS format having units of Jy/pixel. The primary and synthesized beam are then applied to the model spectral line cube, in order to compare the model with ALMA and CARMA submillimeter observations.

\begin{figure*}
\gridline{\fig{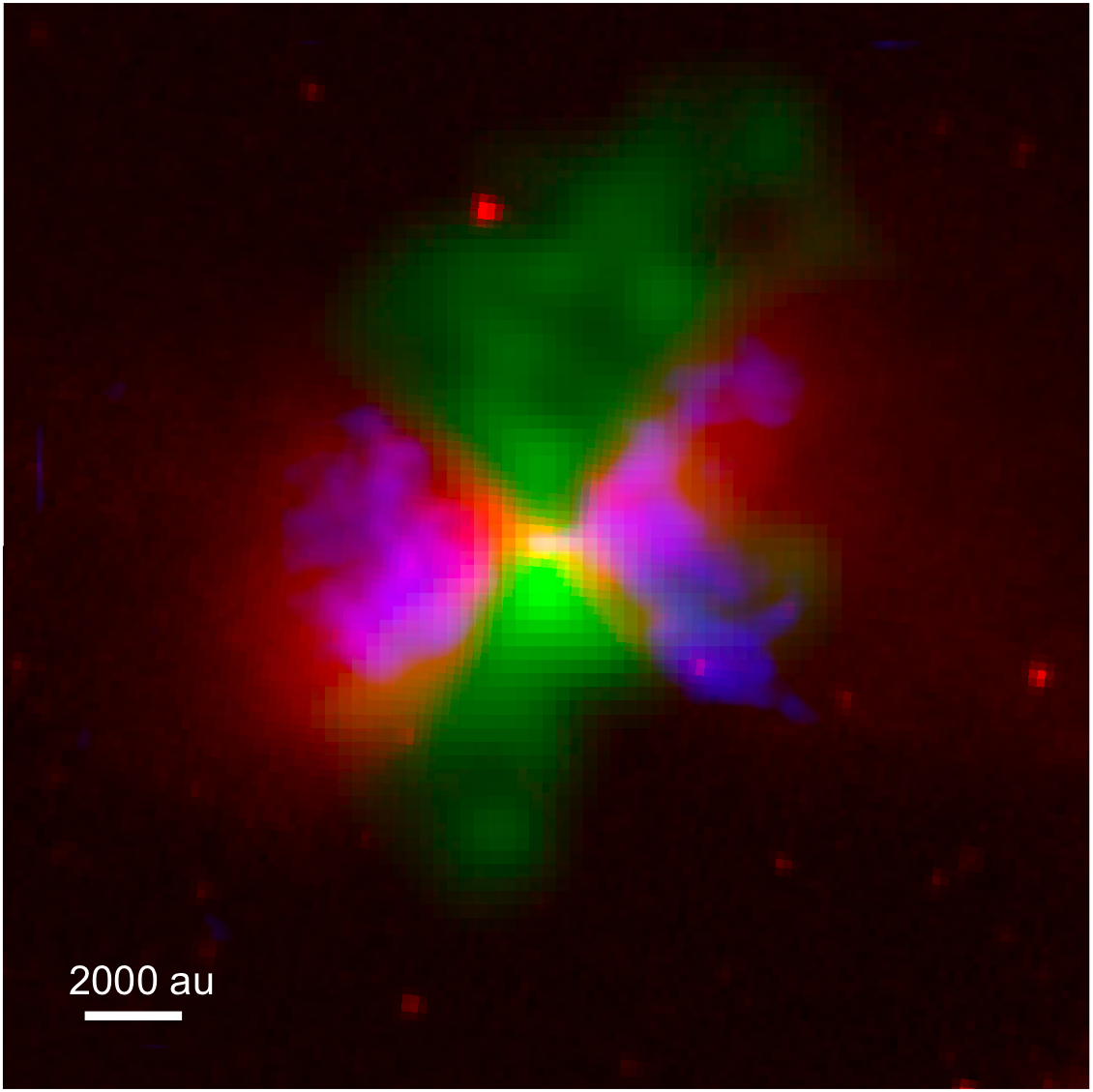}{0.45\textwidth}{(a)}
          \fig{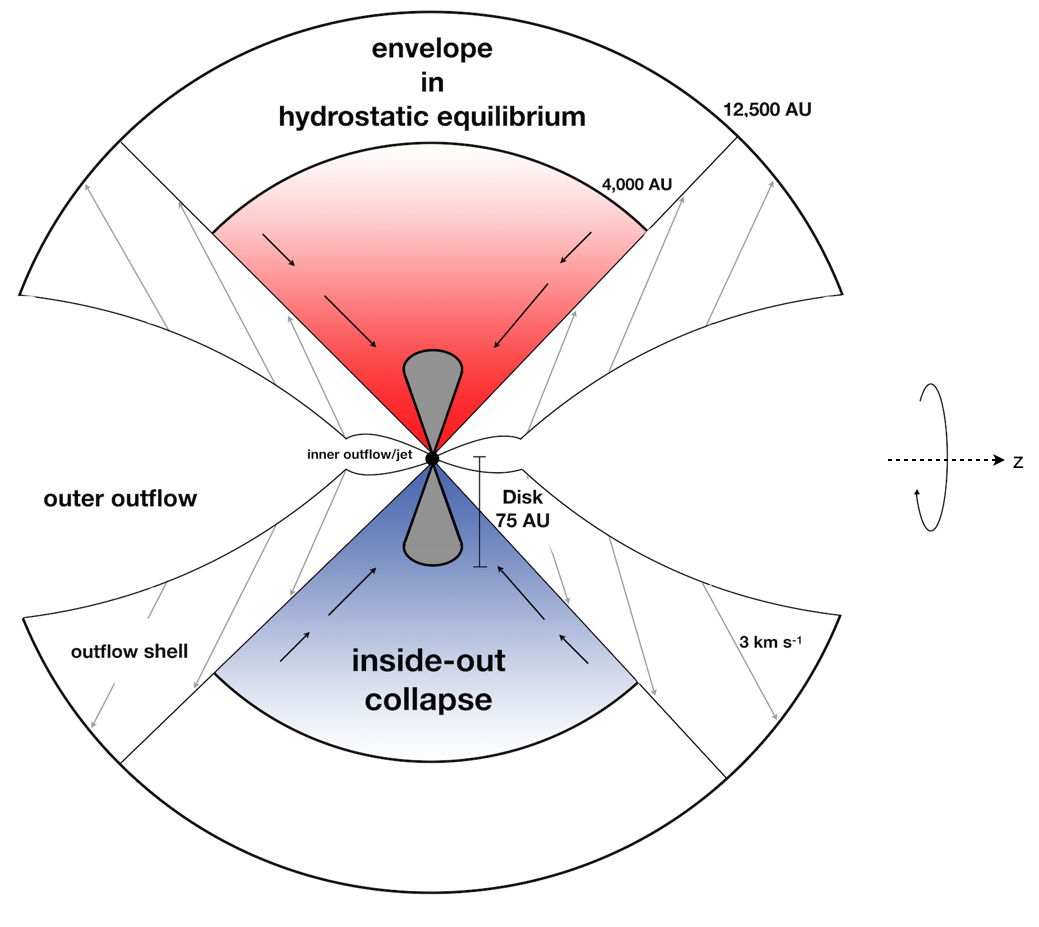}{0.47\textwidth}{(b)}
         }
\caption{$Left$: Three color image of L1527. The north-south emission in green shows dense gas in the envelope as traced by N$_{2}$H$^{+}$ emission from CARMA. The outflow cavity lies east-west. In red, the Spitzer IRAC data at 4.5 $\micron$ traces the outflow cavity in scattered light. Blue reveals the moderate velocity outflow shell as traced by CARMA $^{12}$CO emission. The size of the panel is 160\arcsec$\times$160\arcsec (22,400 au at the distance of L1527). $Right$: Schematic representation of L1527. Regions are not drawn to scale. Each west-east lobe is a low density region with two outflow cavities. The two outflow cavities are surrounded by a denser outflow shell moving at $\sim$3 km~s$^{-1}$ outward, an approximated value obtained by fitting the outflow shell of L1527 with the channel maps of CARMA (See Appendix C). By contrast, in the envelope the velocity increases inwards towards the star due to rotation-infall (collapse) motions. In addition, the direction of the rotation axis of the edge-on disk plus envelope is horizontal (shown in the dashed black arrow), with red-shifted gas found north (top), and blue-shifted gas seen to the south (bottom) of the protostar.} \label{fig:spitzerandschematic} 
\end{figure*}

\subsection{Physical model: Modifications of Hochunk3d}
\label{sec:physicalmodel}

The physical structure of our representative Class 0/I model is based on the \textsc{Hochunk3d} code (\citet{Whitney_2013}). This model uses an MCRT calculation to obtain the temperature of the dust and the gas density profile. The 2D geometry considers the system in spherical coordinates with axial symmetry and a logarithmic increasing grid in radius. The radial domain goes from the dust destruction radius, that is $10.04R_{*}$, to the outer edge of the envelope, $R_{env}=12,500$~au. The meridional domain covers the full 180 degree. The model includes a description of the density structure of the disk component and of the envelope with an outflow cavity. Figure \ref{fig:spitzerandschematic} shows a schematic of the TSC geometry. Both the TSC and UCM models have similar parameters to tune; Table \ref{tab:Physical_Parameters} tabulates adopted values for L1527 to be described more specifically in \S\ref{sec:parameters} and \S\ref{sec:luminosity}.

We adopt a dust model that is appropriate for protostellar environments, based on a comparison of models with observations \citep{Huard_2017}. Specifically we adopt the thinly ice-mantled, coagulated dust of \citet{OH5} (see their Table 1, 5th column), often referred to as “OH5” grains in the literature (e.g., \citet{Evans_2001, Shirley_2005}), augmented by the opacities of \citet{Pollack_1994} at wavelengths shorter than 1.25 $\micron$, as described in \citet{Dunham_2010}. In the next subsections we explain how the main components: envelope (\S\ref{sec:envelope_prescription}), disk(\S\ref{sec:diskenvelope_prescription}), and outflow cavities(\S\ref{sec:outflow_prescription}), are structured in the model and the physical parameters(\S\ref{sec:parameters} \& \S\ref{sec:luminosity}) we chose to model L1527.

\subsubsection{Envelope prescription}
\label{sec:envelope_prescription}

Each region has its own density structure and its distribution profile is semi-analytic. For the UCM collapse model we incorporate the density structure of the envelope as implemented in \textsc{Hochunk3d}. We use a separate calculation to determine the envelope density for the TSC collapse model \citep{Terebey_1984} that consists of an infalling slowly rotating cloud that becomes rotationally flattened as it collapses. These two collapse models are the same in the inner part of the envelope, but differ in the outer envelope due to the fact that UCM neglects gas pressure support. For both collapse models, we set the envelope radius to $R_{env}$ and, for the TSC model, the {\it infall} velocity is zero outside $\sim4000$ au, which is the current boundary of the infall region and it moreover has an infall rate of $3.0\times 10^{-6}$ $M_{\odot}~yr^{-1}$. But for the UCM model, the infall velocity in this outer region is not zero, but is given by the free-fall velocity onto the $0.22~M_{\odot}$ protostar, with a mass infall rate of $5.0\times 10^{-6}$ $M_{\odot}~yr^{-1}$;  moreover, UCM neglects any acceleration from the massive envelope (see \citet{Huard_2017} for additional discussion). Hence the UCM and TSC models presented here have different values for the density of the envelope, but the spatial grid, as well as the outflow and disk definitions, are the same for both models. We expect the difference between the UCM and TSC models to be greatest in the cold outer envelope, where the infall speeds are lowest, less than about five times the thermal line speed (see Table\ref{tab:Physical_Parameters}). This cold gas contributes significantly to the emission near the core of the spectral line, within about 1 km~s$^{-1}$ of line center. The latter depends on the maximum recoverable spatial scale of L1527. The model comparison is meant to show whether the analytic UCM model is sufficient, or whether it is necessary to include the additional complexity of the TSC model in order to match the spectral line profiles.

\subsubsection{Disk-Envelope prescription}
\label{sec:diskenvelope_prescription}

The boundary between the envelope and the disk is not a well explored area, although it could be important during the Class 0/I phase if infalling material goes through a shock when it impacts the disk. In this first paper, we do not include shocks, however, as a first step towards including shocks we modified slightly the shape in which the disk-envelope boundary is defined in \textsc{Hochunk3d} by including a ram pressure boundary condition for motion perpendicular to the disk midplane. The assumption is akin to that of infalling material hitting a ``brick wall" in the disk midplane. Expressing the pressure in terms of the thermal sound speed as $P = \rho c_s^2$, and the ram pressure as $P + \rho v^2$, then the disk-envelope boundary is defined by matching the disk and envelope ram pressures: 

\begin{equation}
\rho_{disk} c_{s_{disk}}^2 = \rho_{env} (c_{s_{env}}^2 + v_{\perp_{env}}^2),  %\label{eq4}
\end{equation}

\noindent
where $v_\perp$ represents the velocity component that is perpendicular to the disk midplane. The thermal sound speed is calculated assuming $c_s = (kT/\mu m_H)^{1/2}$  and assuming that the dust and gas temperatures are equal, which is a reasonable assumption for protostellar densities. The disk boundary that is found using this boundary condition is no longer a flared disk, but has fairly constant opening angle out to the edge of the disk, as described in \S\ref{sec:results}. 

\subsubsection{Outflow cavities and outflow jet prescription}
\label{sec:outflow_prescription}

We use the prescription of \citet{Whitney_2003} for the outflow cavity shape. The shape of both outflow cavities are described by a polynomial function for L1527 which follows $z(\varpi) = \varpi^{b}$ where $\varpi = (x^{2} + y^{2})^{1/2}$ is the cylindrical radius and $b=(inner,outer)=(1.7,1.5)$ is the cavity shape exponent. $\theta_{1}$ and $\theta_{2}$ are the opening angle of the inner and outer cavity surface, respectively, defined at the maximum radius of the envelope. Only the apex of $\theta_{1}$ reach the source center, and $\theta_{2}$ apex starts after 75 au. The density inside the outflow cavities is set to a constant value of $1.6\times 10^{-20}$ g~cm$^{-3}$.

\begin{table*}[t]
\tablenum{1}
\begin{center}
\caption{Physical Parameters.
\label{tab:Physical_Parameters}}
\begin{tabular}{lccccr}
%\multicolumn{3}{c}{\textit{Chemistry model parameters}} \\
\hline \hline
Parameters & Description & TSC & UCM &\\
\hline  
$R_{*} (R_{\odot})$ & Stellar radius & 1.70 & : \\%& Adopted here\\
$T_{*} (K) $ & Stellar temperature & 3,300 & : \\%& Adopted here \\
$M_{*} (M_{\odot})$ & Stellar mass & 0.22 & : \\%& Adopted here$^{*}$\\
$M_{disk} (M_{\odot})$ & Mass of the disk & 0.011 &  0.006 \\
$R_{disk} (au) $ & Disk outer radius & 75 & : \\%& 1,4 \\
$\dot{M}_{disk} (M_{\odot}~yr^{-1})$ & Disk accretion rate & 6.6$\times10^{-7}$ & : \\%& Adopted here \\
$\dot{M}_{env} (M_{\odot}~yr^{-1})$ & Envelope infall rate & 3.0$\times 10^{-6}$ & $5.0\times 10^{-6}$ \\%& Adopted here \\ 
$\theta_{1}(^{\circ}$) & Opening angle of the inner cavity surface & 15 & : \\%& 1\\
z($au$) & z-intercept, inner cavity surface at $\omega$=0  & 75 & : \\%& Adopted here \\
$\theta_{2}$($^{\circ}$) & Opening angle of the outer cavity surface & 6& : \\%& Adopted here \\
$L_{ISRF}(L_{\odot})$ & Luminosity due to ISRF & 0.49 & : \\%& Adopted here \\ 
\hline
\multicolumn{5}{c}{Quantities shown below are derived from input parameters above} \\
\hline
$M_{env} (M_{\odot})$ & Mass of the envelope & 1.77 & 1.04 \\%& Adopted here \\
$c_{s} (km~s^{-1})$ & Thermal sound speed using $\dot{M}_{env} = 0.975 c_{s}^3/G$ & 0.23 & 0.27 \\%&  Adopted here \\
$R_{col} (au)$ & Inside-out collapse radius using 
$R_{col} = c_{s}t_{age}$ & 3800 & n/a \\%& Adopted here \\
$L_{*} (L_{\odot})$ & Stellar luminosity & 0.31 & : \\%& Adopted here \\
$L_{acc,star} (L_{\odot}$) & Stellar hot spot accretion luminosity & 2.14 & : \\%& Adopted here \\ 
$L_{acc,disk} (L_{\odot}$) & Disk accretion luminosity & 0.29& : \\%& Adopted here \\
$L_{int}(L_{\odot})$ & Internal luminosity & 2.74 & : \\ % &  1-4\\

\hline
\multicolumn{2}{@{}l@{}}{\scriptsize The symbol : means the UCM values are the same as TSC values.} \\
\end{tabular}
\end{center}

\end{table*}

\subsubsection{Parameters}
\label{sec:parameters}

Initial values for the parameters were based on the SEDs and image fitting done by Tobin and collaborators \citep{Tobin_2008,Tobin_2010,tobin_2012,Tobin_2013}. The modeling of L1527 is primarily sensitive to $L_{int}$ that depends directly on the chosen values of $R_{*}$ and $T_{*}$ (see Table~\ref{tab:Physical_Parameters}), the radius and temperature of the star, respectively. As described in \citet{Whitney_2013}, to determine the effective temperature of the star, we specify spot parameters such as: two number of spots at 45 degrees in latitude, and 0.10 as the fractional area. Defining $R_{*}$ and $T_{*}$, we keep the internal luminosity $L_{int} = 2.74~L_{\odot}$ fixed which is reasonable given the range of possible luminosity values for L1527. For a Class 0/I protostar, the protostar mass should be small. Therefore we looked at lower mass values in the literature, and selected a central star mass of $M_{*}$=$0.22M_{\odot}$, a mass that best fit the position-velocity diagram for $^{13}$CO \citep{tobin_2012}. The disk outer radius is the centrifugal radius that is set to 75 AU \citep{Aso_2017}. The disk inner radius is the dust destruction radius, a value that is set by the code depending on the stellar luminosity.

Two important considerations are crucial to matching both of the SED apertures, and for both TSC and UCM for our modeled L1527. Using identical parameters leads to somewhat different looking SEDs; however we find that solely modifying $\dot{M}_{env}$ and $M_{disk}$ brings the SEDs for UCM and TSC into reasonable correspondence, especially near 100 $\micron$ wavelength, near the far-infrared peak of the SED. In this case, the reasonable parameters found are $M_{disk}$=0.006$M_{\odot}$ and a $\dot{M}_{env}$ of 5.0$\times 10^{-6}$ $M_{\odot}~yr^{-1}$ for the UCM model. For the TSC model, on the other hand, the mass of the disk chosen is 0.011$M_{\odot}$ and a $\dot{M}_{env}$ of 3.0$\times 10^{-6}$ $M_{\odot}~yr^{-1}$. The second consideration is the outflow; from newer data we find there is evidence in both the CARMA and ALMA $^{12}$CO data for a narrow jet, plus a wide-angle bipolar outflow, with a transition that happens at $\sim 75$ au from the protostar. The ALMA data (Fig.\ref{fig:coalma}) show the inner region, including the jet and outflow lobes in a representative velocity channel. The CARMA $^{12}$CO data in Fig.\ref{fig:spitzercarma} and Fig.\ref{fig:cochannels} channel maps also display a narrow outflow structure near the protostar. We model this structure (Fig.\ref{fig:tempdens}) using a dual outflow cavity with the narrow ``jet'' extending to 75~au, a radius that is similar to the 85~au suggested by a cavity modeling analysis from \citet{Tobin_2010}. We also specify a small, but nonzero density in the outflow cavity ($1.6\times 10^{-20}$ g~cm$^{-3}$); adding a small amount of material (having standard dust-to-gas mass ratio of 0.01) to the outflow cavity produced a better fit to the SED from that shown in \citet{Tobin_2010}. Our choice of outflow parameters removes the need for the puffed up disk required by previous modeling \citep{Tobin_2010}. For the purposes of this paper we did not attempt a detailed modeling of the jet length or the cavity shape. 

\subsubsection{Luminosity}
\label{sec:luminosity}

The internal luminosity is fixed at $L_{int} = 2.74~L_{\odot}$ to match the 2.74 $L_{\odot}$ internal luminosity found by \citet{Tobin_2008}.  The internal luminosity includes contributions from the star $L_{*}$, plus the accretion luminosity from material falling onto the star $L_{acc,star}$, and the accretion luminosity generated within the disk $L_{acc,disk}$, as described in \citet{Whitney_2013}. For a Class 0/I protostar, the protostar mass should be small and most of the luminosity should be due to accretion. This led to choosing a stellar radius of 1.70$R_{\odot}$ with an effective temperature of 3,300 K that gives a stellar luminosity of 0.31 $L_{\odot}$ (see Table \ref{tab:Physical_Parameters}). One additional term $L_{ISRF}$  contributes to the total luminosity of the system, $L_{tot}=L_{int} + L_{ISRF} =3.23~L_{\odot}$; the term $L_{ISRF}$ is due to external illumination by the Interstellar Radiation Field (ISRF) and is based on the galactic value computed near our solar system; see description in \citet{Huard_2017}. 

The largest luminosity term is due to the stellar hot spot accretion luminosity, which is defined as:

\begin{center}
\begin{equation}
L_{acc,star} = GM_{*}\dot{M}_{disk}(\frac{1}{R_{*}}-\frac{1}{R_{trunc}})
\label{eq8}
\end{equation}
\end{center}

where $R_{trunc}$ is the truncation radius where the disk is truncated by the stellar magnetospheric field and its value is approximately the same as in \citet{Tobin_2008}. The disk accretion rate $\dot{M}_{disk}$ is an important parameter that leads to the stellar accretion luminosity that encompasses about 66$\%$ of the total luminosity of the system. The code determines that the value  $\dot{M}_{disk}$=6.6$\times$10$^{-7}$ $M_{\odot}$~yr$^{-1}$ leads to an accretion luminosity onto the star of 2.14 $L_{\odot}$.

We also include the disk accretion luminosity in terms of the energy dissipated at the inner boundary of the disk \citep[see][for more details]{shakura_1973,lynden_1974,kenyon_1987, Whitney_2003_I}. Table \ref{tab:Physical_Parameters} summarizes the values of the different luminosity terms.

\begin{deluxetable*}{lclclcll}
\tablenum{2}
\tablewidth{0pt}
\tablecaption{\label{tab:frac} Isotopic fractionation reactions used in the model. $\Delta$E values are taken from \citep{Langer_1984}. $\Delta$E values for reactions involving $^{17}$O are assumed to be the same as for the equivalent reaction of $^{18}$O, with the pre-exponential factor in the rate calculation scaled by the reduced mass \citep{Young_2007}. }
\tablehead{\colhead{Reactions}}
\startdata
$^{13}$C$^+ $& + & CO & $\rightleftharpoons$ & C$^{+}$ & + &  $^{13}$CO & $\Delta$E=35K\\
$^{13}$C$^+$ & + & C$^{18}$O & $\rightleftharpoons$ & C$^{+}$ & + & $^{13}$C$^{18}$O & $\Delta$E=36K\\
HCO$^+$ &  + & $^{12}$CO & $\rightleftharpoons$ & CO & + & H$^{13}$CO$^+$ & $\Delta$E=9K  \\
HCO$^+$ &  + & C$^{18}$O & $\rightleftharpoons$ & HC$^{18}$O & + & CO & $\Delta$E=14K  \\
HCO$^+$ & + &  $^{13}$C$^{18}$O & $\rightleftharpoons$ & H$^{13}$C$^{18}$O$^+$ &  +  & CO & $\Delta$E=22K \\
H$^{13}$CO$^+$ & + & C$^{18}$O & $\rightleftharpoons$ & HC$^{18}$O$^+$ & + &  $^{13}$CO & $\Delta$E = 5K  \\
H$^{13}$CO$^+$ & + &  $^{13}$C$^{18}$O & $\rightleftharpoons$ & H$^{13}$C$^{18}$O$^+$ & + &  $^{13}$CO & $\Delta$E=13K \\
HC$^{18}$O$^+$ & + &  $^{13}$C$^{18}$O & $\rightleftharpoons$ & H$^{13}$C$^{18}$O$^+$ & + &  C$^{18}$O &  %$\Delta$E=8K \\
\enddata
\end{deluxetable*}

\subsection{Chemical model}
\label{sec:chemicalmodel}

We carry out local chemical evolution modeling to determine the abundances of the molecules that have been observed.  The chemical models are time-dependent, and we focus on the results at two epochs 10$^{4}$ and 10$^{5}$ years that span the range of possible ages for L1527.

Our chemical network is taken from the UMIST database, RATE12 \citep{Umist_2013}. The reactions of the carbon and oxygen isotopes have been added such that each reaction involving an atom of the major isotopes will have an equivalent reaction involving the minor isotopes \citep[see][for more details]{Willacy_2009_II}.  Fractionation of the oxygen and carbon isotopes can occur via the reactions listed in Table~\ref{tab:frac}. For reactions involving $^{17}$O the values of $\Delta$E are taken to be the same as the equivalent reaction of $^{18}$O, and the pre-exponential part of the rate calculation is scaled by the reduced mass \citep{Young_2007}.

The network also includes gas-grain reactions, i.e.\ freezeout on the grain \citep{Hasegawa_1993}, thermal desorption \citep{Hasegawa_1992}, as well as photodesorption, and desorption by heating of grains by cosmic rays \citep{Oberg_2009a,Oberg_2009c}. The freezeout and desorption reactions are also described in \citet{ww09}. Freezeout is assumed to occur with a sticking coefficient of 1.0 \citep{Bisschop_2006} for all species. For desorption processes, the binding energies required are taken from UMIST12. Cosmic ray heating rates are given by 
\begin{equation}
    k_{crh} = 3.16\times10^{-19} \times \exp(-E_b/70.)
\end{equation}
\citep{Hasegawa_1993}, where $E_b$ is the binding energy of the accreted molecule.  Photodesorption rates are given by 
\begin{equation}
    k_{photd} = F Y <\pi a^2 n_g> \Theta_X
\end{equation}
\citep{wl00} where F is the UV field (the total of the stellar, interstellar and cosmic ray induced fields in units of $G_{0}$), Y is the yield per photon which is taken to be 10$^{-3}$ for all species \citep{Westley_1995}, except for oxygen atoms (Y=10$^{-4}$) and H$_2$O (Y=10$^{-3}$) and Y = 2 $\times$ 10$^{-3}$ for desorption as OH \citep{Hollenbach_2009}. The dust is assumed to be well mixed with the gas, $n_g$ is the number density of dust grains ($n_g$ =10$^{-12}$ $n_H$ ), and the average $< \pi a^{2} n_g>$ = 2.1 $\times$ 10$^{-21}$ $n_H$ (standard interstellar value). $\Theta_X$ is the surface coverage of species $X$ = ($n_s(x)$/$\Sigma n_s(y)$), where $n_s(x)$ is the abundance of X in the ices and $\Sigma n_s(y)$ is the total abundance of ices). The stellar UV field is assumed to have a typical T Tauri value of 500 $G_0$ at 100 au (unextincted) from the star, where $G_0$ is the standard ISRF \citep{Bergin_2003}. The dense envelope generates significant extinction. To account for this, the local UV field is decreased to take into account the extinction calculated along the line of sight of the star.

Grain surface reactions are included using the approach of \cite{Garrod_2011}.  Only atoms, H$_2$ and simple hydrides (OH, CH, NH, and their isotopologues) are assumed to be mobile on the grain surface. 

Initially we assume all elements are in their atomic form, except for carbon which is ionic and hydrogen which is 95\% molecular. The initial abundances used are given in Table~\ref{tab:initabu}. We assume a cosmic ray ionization rate of $\zeta_{CR}$ = 1.3 $\times$ 10$^{-17}$ s$^{-1}$.

For the photodissociation of CO (and its isotopologues), and H$_{2}$ we use the self-shielding coefficients provided by \citet{Visser_2009} assuming a doppler width of 0.3 kms$^{-1}$ and isotopic ratios of $^{12}$C/$^{13}$C = 89, $^{16}$O/$^{18}$O = 498 and $^{16}$O/$^{17}$O = 1988, which are taken to be the same as local ISM values \citep{refId0}. Abundances here are prescribed as $X$, the fractional abundances relative to total hydrogen, $n_{X}/(n_{H}+2n_{H{_2}})$.  

Starting from $t=0$, the molecular abundances grow rapidly until $t=10^{4}$ years, when the abundance of $^{12}$CO reaches its maximum achievable value of $7.22\times10^{-5}$, a value that is set by the assumed carbon abundance (Table \ref{tab:initabu}) and that is based on Taurus observations. Therefore we chose $t=10^{4}$ yrs as our starting reference time for the chemistry. The grid in the chemical model follow the same structure as in the physical model (\S\ref{sec:physicalmodel}) but, in order to speed up the computation, the chemistry was only computed on every 5th grid point in the spatial grid. The solution of the chemical reaction network reaches good convergence throughout most of the spatial grid, including the disk, envelope, and outflow shell. However, we exclude from consideration the low density ($n$=4000~cm$^{-3}$) outflow cavity due to reduced convergence in this mostly atomic region. There is little impact on our study because the outflow cavity contributes little to the molecular emission that is the focus of this investigation. However, we do capture the chemistry that happens based on our prescription of the outflow shell (see Fig.\ref{fig:cochannels}) when we add constant velocity in this region between the envelope and outflow, see \S2.4 for more details.

\begin{center}
\begin{deluxetable*}{ccc}
\tablenum{3}
\tablecaption{CARMA observational parameters.\label{tab:carma_obs_parameters}}
\tablewidth{0pt}
\tablehead{
\colhead{} & \colhead{CARMA} \\
\colhead{} & \colhead{$^{12}$CO(1-0)} & \colhead{N$_{2}$H$^{+}$(1-0) }\\
\colhead{Target, date} & \colhead{L1527 IRS, August and November 2009}
}
\startdata
Coordinate Center & R.A.(J2000)=4$^\mathrm{h}$39$^\mathrm{m}$53$^\mathrm{s}$.9000 & \\
& Dec.(J2000)=26$^\mathrm{\circ}$03$^\mathrm{'}$10$^\mathrm{"}.000$\\
Frequency & 115.271 GHz & 93.17378 GHz  \\
Synthesized beam & 3.32$\arcsec$$\times$2.95$\arcsec$ & 10.97$\arcsec$$\times$8.73$\arcsec$ \\
Primary beam & 54$\arcsec$ & 67$\arcsec$\\
Velocity resolution & 0.34 km s$^{-1}$ & 0.26 km s$^{-1}$ \\
Noise level (detected channel) & 0.187Jy beam$^{-1}$ & 0.200Jy beam$^{-1}$ \\
\enddata
\end{deluxetable*}
\end{center}

\subsection{Synthetic line images}
\label{sec:modelvisualization}

Step 3 of RadChemT calculates the protostellar environment in a PV cube. Because \textsc{Hochunk3d} did not include spectral line emission we selected \textsc{RADMC-3D}, version 0.41 \citep{dullemond_2012}
,
to generate synthetic spectral line emission assuming LTE. LTE improves computational speed and holds in locations where the gas density is above the critical density. The generation of synthetic line images takes into consideration the density, temperature, and velocity grids from Step 1, and the abundance grid from Step 2 at $t = 10^4$ year and $t = 10^5$ year time stamps, all of which are translated into the \textsc{RADMC-3D} file format. 

Finally, for each model density distribution and epoch, we carry out line-of-sight radiative transfer calculations using \textsc{RADMC-3D} to construct synthetic line and continuum observations, and compare these against the multi-telescope data set for L1527. 

Additional inputs for L1527 in RADMC-3D were source inclination $i = 85\degr$, distance $d = 140~pc$, and $v_{sys}$ = 6.0~km~s$^{-1}$ system velocity. For consistency the OH5 dust opacity law is the same as that used for the step 1 continuum radiative transfer. Standard and reasonable assumptions for the protostellar environment are $T_{dust}=T_{gas}$ and 100 for the gas-to-dust ratio. The microturbulent velocity was fixed at 0.1~km~s$^{-1}$, a value that is required for LTE and results in smoothing the line profile. A larger microturbulent velocity value could affect the hyperfine structures of the molecular spectrum by causing line blending. However, data from CARMA CO observations indicate somewhat larger values for the microturbulence should be used for the outflow shell region. Based on the $^{12}$CO data presented in \S\ref{sec:n2h+_carma}, and in Appendix C, we chose a very simple outflow velocity prescription having a constant outward radial motion of 3~km~s$^{-1}$ in the outflow cavity and also in the outflow shell. However a careful treatment of the outflow shell lies outside the scope of this paper. From the inputs \textsc{RADMC-3D} creates a synthetic model PV cube in FITS format having units of Jy~pixel$^{-1}$.

Comparison with observations also requires applying telescope specific parameters. The velocity channel width and spatial pixel size were specified as inputs to \textsc{RADMC-3D}. The synthetic images are sampled in real space in order to be able to see more clearly the variations in the chemical abundances and, therefore, the dynamical range of the system, in which in our models are greater than the interferometric data. To compare with millimeter interferometer data, the synthetic model images are multiplied by a Gaussian primary beam (peak value normalized to unity), and then convolved with a circular Gaussian synthesized beam (area normalized to unity). See \S\ref{sec:observations} for ALMA and CARMA observational parameters.

\section{Observations} 
\label{sec:observations}

\subsection{CARMA data}

L1527 was observed on August and November 2009 using the Combined Array for Research in Millimeter-wave Astronomy (CARMA) located at an altitude of 7200 feet on the Eastern California Inyo mountains. Observations were obtained in D- and C-array configurations, which provide an uniform uv-coverage between 9 and 371 m. The CARMA correlator was set to observe the $^{12}$CO (1-0) emission line ($\nu$ = 115.271 GHz) in the upper side band, and the $^{13}$CO (1-0) ($\nu$ = 110.201 GHz) line in the lower side band. These two lines were observed at a velocity resolution of 0.34 km s$^{-1}$ in two 8 MHz spectral windows. The dust continuum emission was observed in two 1 GHz windows separated by 3.6 GHz and centered at the mean frequency of 112.73625 GHz ($\lambda$ = 2.66 mm). Here we do not use $^{13}$CO (1-0) since it has higher optical depth than C$^{18}$O, therefore, making it more difficult to see the emission from the disk. The band pass shape was calibrated by observing 3C84 and the flux calibration was set by observing Uranus. The quasar 3C11 was observed every 12 minutes to correct for atmospheric and instrumental effects. The data reduction and the image reconstruction were obtained using the MIRIAD software package. For N$_{2}$H$^{+}$ data and model comparison, we convolve the model using a geometric mean FWHM of 9$\arcsec$, namely, $b=\sqrt{(b_{max}*b_{min})}=\sqrt{10.97\arcsec*8.73\arcsec} = 9\arcsec$. The observational parameters are described in Table.\ref{tab:carma_obs_parameters}.

\subsection{ALMA data}

The observational data presented for C$^{18}$O (2-1) and having ${0.96\arcsec \times 0.73\arcsec}$ (that is the geometric beam) spatial resolution and a maximum recoverable scale of $\sim15\arcsec$ are based on data from the ALMA archive that was taken during cycle 0 on 2012 August 26 (Project code: 2011.0.00210.S). Higher spatial resolution ALMA data exist: however they resolve out much of the $\sim 10\arcsec$ emission that is relevant to this study. We convolve the model using the same geometric mean formulation as in CARMA, giving $b = 0.8\arcsec$ for the ALMA spatial resolution.  The model is also multiplied by the ALMA primary beam, which was taken to be a smooth $28\arcsec$ FWHM Gaussian image. Table~\ref{tab:alma_obs_parameters} summarizes observational parameters. More information about the observations and calibration is given in Table 1. in \citet{Ohashi_2014}. 

\section{Results and discussions} \label{sec:results}

\subsection{L1527 SED fits}

\begin{figure*}
\gridline{\fig{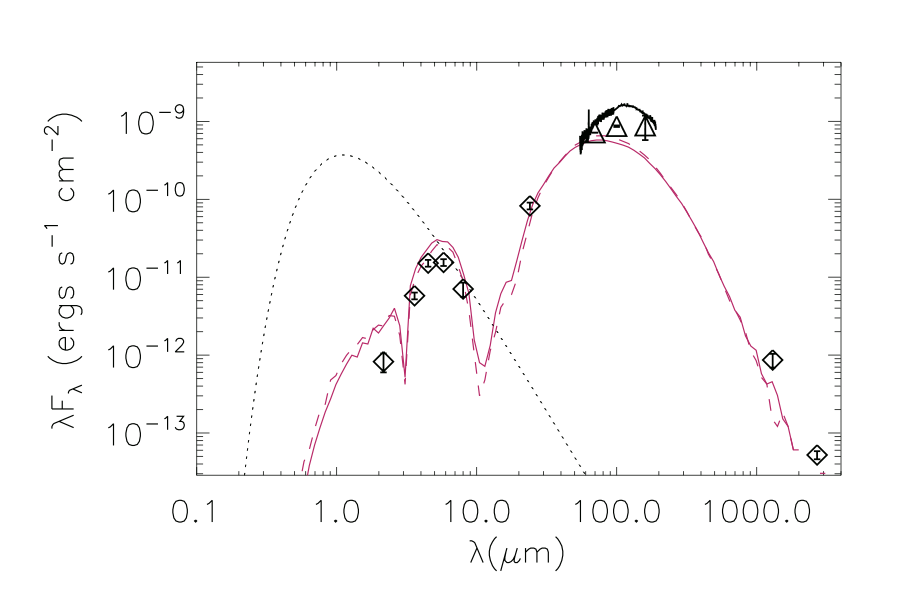}{0.5\textwidth}{(a)}         
          \fig{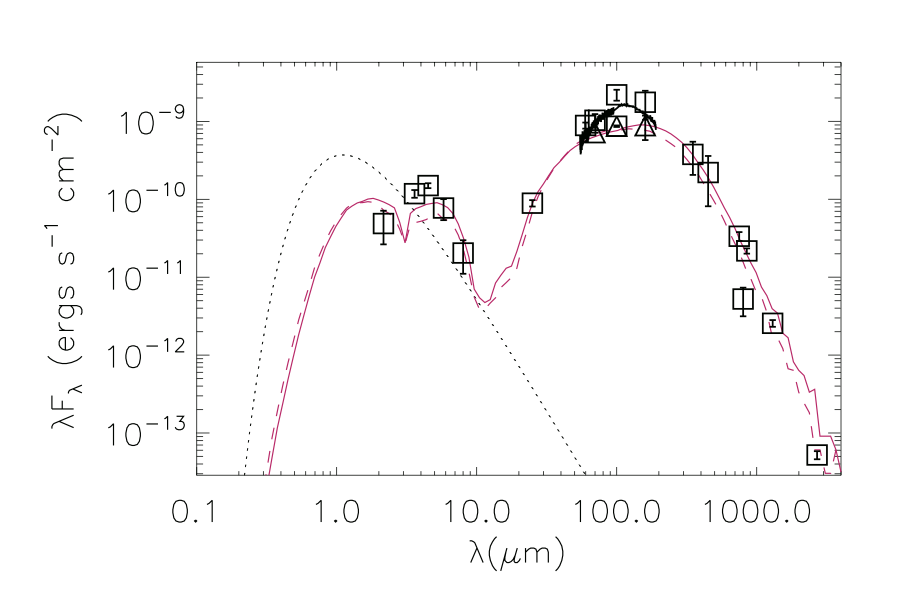}{0.5\textwidth}{(b)}
          } 
\gridline{\fig{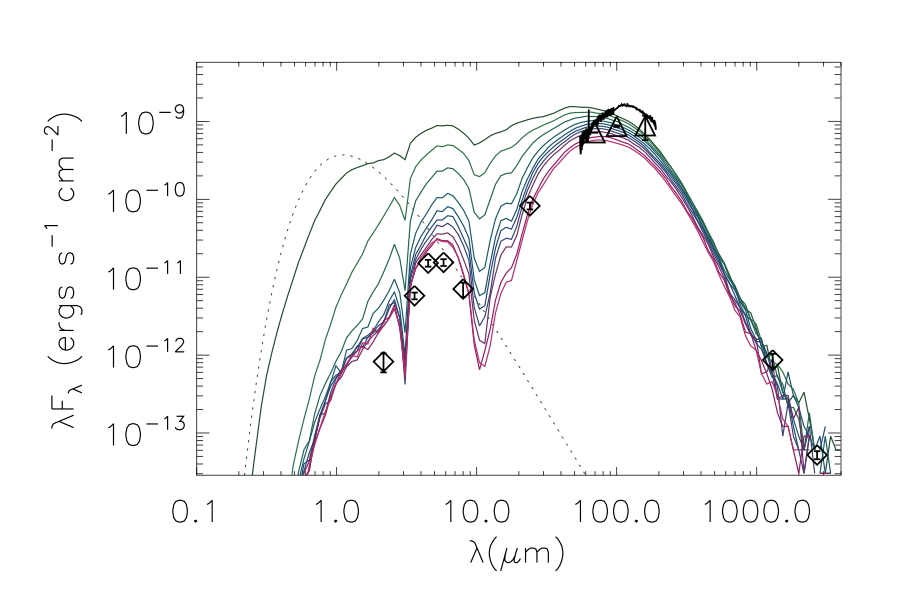}{0.5\textwidth}{(c)} 
          }
\caption{Spectral Energy Distribution (SEDs) of L1527 is plotted at 85$^{\circ}$ source inclination for both the UCM (dashed pink line) and TSC (solid pink line) collapse models. Panel (a) and Panel (b) show the flux density data obtained from \citet{Tobin_2008} as diamond icons for a model aperture size of 1000 au ($7.14\arcsec$) and as squared icons for a model aperture size of 10,000 au ($71.4\arcsec$), respectively. Except; the \textit{Herschel} CDF spectrum (solid black line) and HPPSC data (triangles) plotted near 100~\micron\ are the same in all three panels. See Appendix B, Table \ref{tab:photometry_7.14} and Table \ref{tab:photometry_71.4} for a detailed description. To illustrate the effect of differing source inclination, panel (c), in particular, shows 10 inclinations (pole-on green, edge-on pink).}
\label{fig:sed}
\end{figure*}

We present a comparison of the flux density of the simulated L1527 for both UCM and TSC models. The simulated L1527 SEDs are shown in Figure \ref{fig:sed} plotted at 85$^{\circ}$ edge-on (pink), the appropriate inclination of L1527 \citep{Oya_2015}. Panel (a) and (b) show the model SEDs computed for a 1000~au (7.14$\arcsec$) aperture and a larger 10,000~au (71.4$\arcsec$) aperture, respectively. We also include a plot to illustrate the strong effect of source inclination on the SED model curves. We re-generate SEDs for L1527 by including the flux density values from \citep{Tobin_2008} plus we added flux density values from \textit{Herschel} (see Table \ref{tab:photometry_7.14} and \ref{tab:photometry_71.4}), that come mostly from the thermal radiation of the envelope. The \textit{Herschel} data were downloaded from the \textsc{IRSA/IPAC} archive. Since the \textit{Herschel} CDF spectrum \citep{Green_2016} and HPPSC catalog flux density points \citep{marton2017herschelpacs} have apertures of $6 - 14\arcsec$, namely that lie in between the model 7.14$\arcsec$ and 71.4$\arcsec$ apertures, we chose to show the \textit{Herschel} data on both SED aperture plots. Although we do not intend to redo the best parameter fit for L1527, here we revisit the outskirts of the envelope where the emission is also important for the chemistry.

Overall, both the UCM and TSC models provide reasonable SED fits, with the TSC model providing a better fit for the 10,000~au aperture that is consistent with the large spatial extent of the system. Another regime that supports our later statement is the region where the disk emits, this is between 2.16 and 8.0 $\micron$. We find that the TSC is a closer fit to the flux points although not perfectly due the different dust opacity population and density that lies in here which not necessarily is in accordance with the dust opacity model we use for the envelope. The PACS data are also of particular interest because they occur near the peak of the SED distribution. The \textit{Herschel} PACS HPPSC data have an aperture of 6 \arcsec at 70 $\micron$ and 100 $\micron$, and 12\arcsec\ at 160 $\micron$. The PACS data at 70 $\micron$ and 100 $\micron$ have a spatial resolution that is comparable with the 7.14 \arcsec aperture model (Fig.\ref{fig:sed}, Panel(a)) and moreover, the flux density values do not deviate much from the modeled ones. However, for PACS $>$100 $\micron$ it is slightly offset since the discrepancy in aperture size is greater. The \textit{Herschel} spectrum does not match the \textit{Herschel} photometry points either, so the \textit{Herschel} data are not consistent with each other at around 100 $\micron$ and longer wavelengths. Some possible explanations can be that: a) the source is extended, which might mean the data calibration is off/incorrect (data issue) or b) the source is varying in luminosity (source variability). 

The properties of the central region were chosen carefully (see \S\ref{sec:parameters})  to best describe the physical structure of L1527 as constrained by imaging data and the SED photometry.  We estimated these stellar properties properly describing L1527 based on \citet{Tobin_2008, Tobin_2010, Tobin_2013}. We also adjusted the description of the outflow cavity to transition from jet to wide-angle outflow based on ALMA CO data (Fig.\ref{fig:coalma}) at 75~au (see \S\ref{sec:parameters}).  For the fixed protostellar parameters listed in Table \ref{tab:Physical_Parameters} and described in \S\ref{sec:parameters} and \S\ref{sec:luminosity}, we varied the mass infall rate and disk mass, finding the mass infall rate, $\dot{M}_{env}$, to be 5$\times$10$^{-6}$ $M_{\odot}$~yr$^{-1}$ and 3$\times$10$^{-6}$ $M_{\odot}$~yr$^{-1}$ for the UCM and TSC models, respectively. Based on the magnitude of $\dot{M}_{env}$ chosen, the age of the simulated L1527 is estimated to be $\frac{0.22M_{\odot}}{3\times10^{-6}M_{\odot}~yr^{-1}}$ $\sim$ 7$\times$10$^{4}$ yrs. We adopt that L1527 is in the protostar collapse phase with an age of $t\simeq$10$^{5}$ yrs.

To further assess the self-consistency of the model, we note that in every evolutionary stage of the star formation process, the mass budget of the infalling envelope and accreting disk have to be consistent with the feedback of the outflows and winds. In the absence of any outflow, then  $\dot{M}_{env} = \dot{M}_{disk}$, all of which falls onto the central star. As discussed in section \ref{sec:luminosity} and shown by Equation (\ref{eq8}), the disk accretion $\dot{M}_{disk}$ leads to generous $L_{acc,star}$ accretion luminosity, which is fit by the modeling procedure. Therefore a mismatch in the two accretion rates is related to outflow feedback. The infall efficiency is given by the ratio of the accretion rates; for L1527 our model value of  $\frac{\dot{M}_{disk}}{\dot{M}_{env}}$ $=6.6\times10^{-7}M_{\odot}~yr^{-1}/ {3\times10^{-6}M_{\odot}~yr^{-1}}= 0.22$. 
We point out that this value is similar to the value of 0.25 estimated for the protostar TMC-1 \citep{Terebey_2006}. These values differ from unity, and provide interesting constraints to theoretical discussions of star formation efficiency. See \citet{Matzner_2000}, and in the context of high mass star formation, \citet{Zhang_2014} for an extensive treatment of outflow feedback, and its relation to star formation efficiency.

Next, for comparison purposes with \citet{van_t_Hoff_2018}, we produced a second UCM run by increasing the stellar mass to 0.45 $M_{\odot}$ and the disk radius to 125 au. The untuned model produced SED fits that were worse but still reasonable. No significant difference could be discerned in the density, temperature, or resultant chemistry. However, the velocities increased by $\sqrt{2}$ due to higher gravity from the larger mass. The effect was clearly seen in model spectral line profiles and PV diagrams (see further in \S\ref{sec:c18o_alma}), therefore, turning it into a promising venue to investigate the dynamical mass of L1527 in future studies.

\subsection{Temperature and density maps}

\begin{figure*}

\gridline{\fig{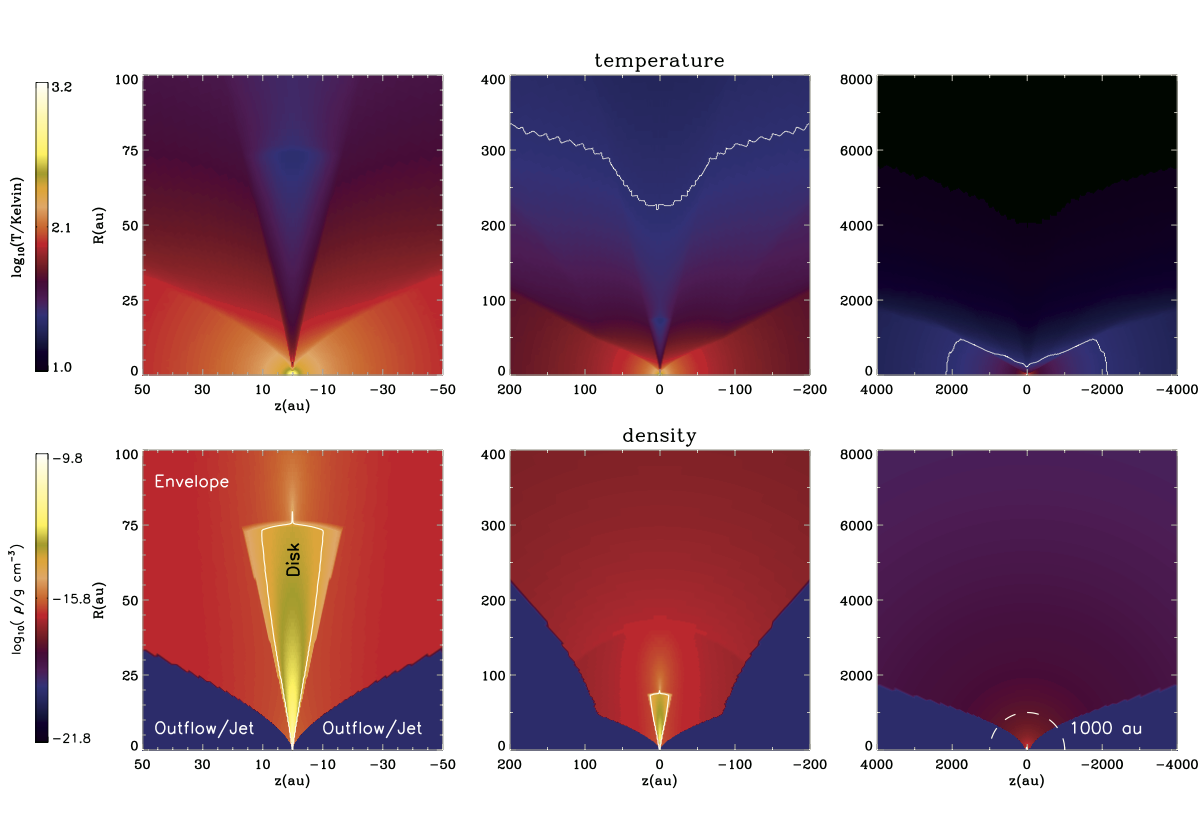}{1.0\textwidth}{}}

\caption{Meridional ($z$ vs. cylindrical $r$) temperature (top) and gas density (bottom) panels for the disk + envelope + outflow using TSC model of L1527. The edge-on disk is oriented vertically in the figure, and rotated by 90$^\circ$ with respect to the outflow cavity. The shape of the dual outflow cavity is evident in the lower middle panel in blue. \textit{White line contours:} T = 25K (top panels) and $n$ = 10$^{9}$ cm$^{-3}$ (first two bottom panels). The white line contour in the third bottom panel shows the radius for the 1,000 au aperture size. The colorbar temperature wedge is $\log_{10}$($T$/Kelvin), and the density wedge is $\log_{10}$($\rho$) where $\rho$ is the mass density in g~cm$^{-3}$.\label{fig:tempdens}}
\end{figure*}

Figure \ref{fig:tempdens} presents the dust temperature (top panels) and gas density (bottom panels), computed using the RadChemT package, as meridional cuts through the envelope + disk + bipolar outflow cavity. Three different zoomed views are shown from left to right. The central star is located at the origin, the outflow cavities extend horizontally, and the edge-on disk (vertical) is oriented 90$^\circ$ with respect to the outflow cavities. The dust temperature is calculated from the radiative equilibrium solution using \textsc{Hochunk3d} and its distribution goes as the power $r^{-0.5}$ in optically thick regions, and as the power $r^{-0.33}$ in optically thin regions \citep{Kenyon_1993}. Near the protostar, the temperature reaches the sublimation temperature of $\sim$1600 K inside the outflow cavities, and the small amount of dust sublimates due to exposure from stellar radiation in this region, making it feasible for the atomic gas to be present at vibrational energy levels. The radius of the collapsing region (i.e. expansion wave) is $3,800$~au, and grows larger in time at the sound speed. Outside the collapse radius, the distribution of the envelope is a power-law function of the radius $\rho \sim r^{-2}$ . Inside the collapse radius the density distribution becomes flat, transitioning at smaller radius to the free-fall zone where the density behaves as $\rho \sim r^{-3/2}$ well outside the disk \citep{Shu_1977, Terebey_1984}. Contour lines in white show $T = 25$ K (top panels), our definition of the evaporation temperature of CO(e.g, \citet{2015ApJ...813..128Q,2019ApJ...882..160Q,2019MNRAS.485.1843W}, and number density $n= 10^9$ cm$^{-3}$ (bottom panels). The shape of the dual outflow cavity is seen most clearly in the lower middle density panel in blue; near $75$~au the narrow outflow/jet opens into a wide-angle outflow. The circle in the lower right shows the (small) $1000$~au radius aperture ($7.14\arcsec$) that is used for aperture photometry. Note that this aperture covers a small portion of L1527 making necessary to compare PACS data points with 71.4$\arcsec$ ($10,000$~au) that covers a much larger region. 

The outer disk radius equals the centrifugal radius of the envelope, 75 au, and the disk-envelope boundary is wedge-shaped due to the ram pressure boundary condition (see \S\ref{sec:physicalmodel}).  The highest density region is found at the disk midplane with the number density on the order of $\sim$10$^{14}$ cm$^{-3}$. At this high density region inside 50 au, the temperature in the disk midplane is slightly higher than 30 K. Beyond 75 au, the midplane is shielded from the stellar radiation leading molecules to gradually freeze-out onto dust grains until $T = 25$ K at $\sim$225 au, where the CO snowline lies and freeze-out is expected to happen more rapidly. Beyond the $T = 25$ K contour line, the temperature decreases until reaching the typical temperature of the envelope, $T= 10$ K. Since we do not include shocks, there is no increase in the temperature profile due to them. The luminosity that shocks can produce is a small fraction compared to the stellar luminosity, however, it is been suggested that can be enough to liberate molecules from grains \citep{Sakai_2014}. Comparing the UCM and TSC models, the temperature and the density distribution from both show very little difference in terms of structure.  

\begin{figure*}

\gridline{\fig{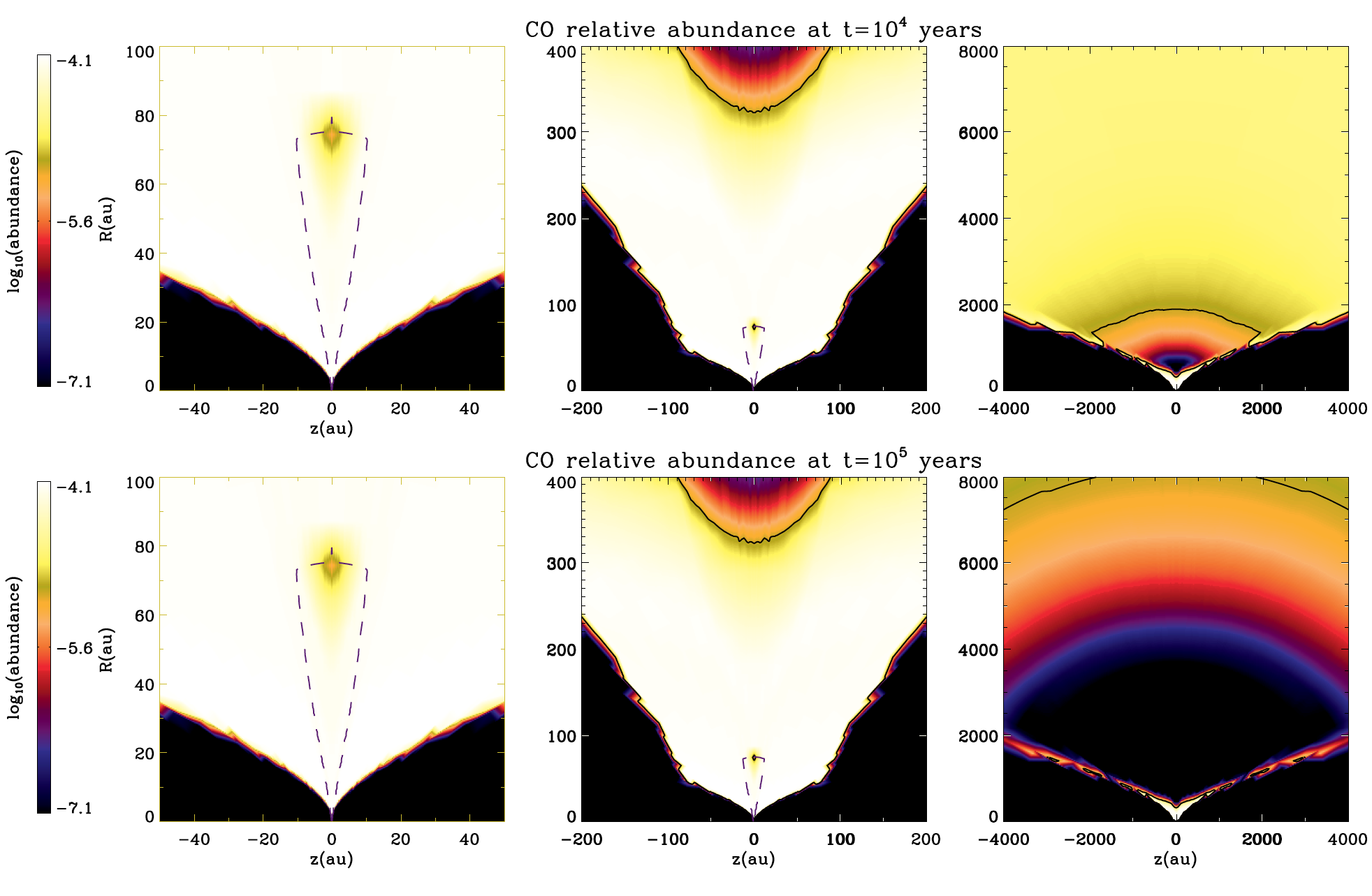}{0.8\textwidth}{(a)}
         }
\gridline{\fig{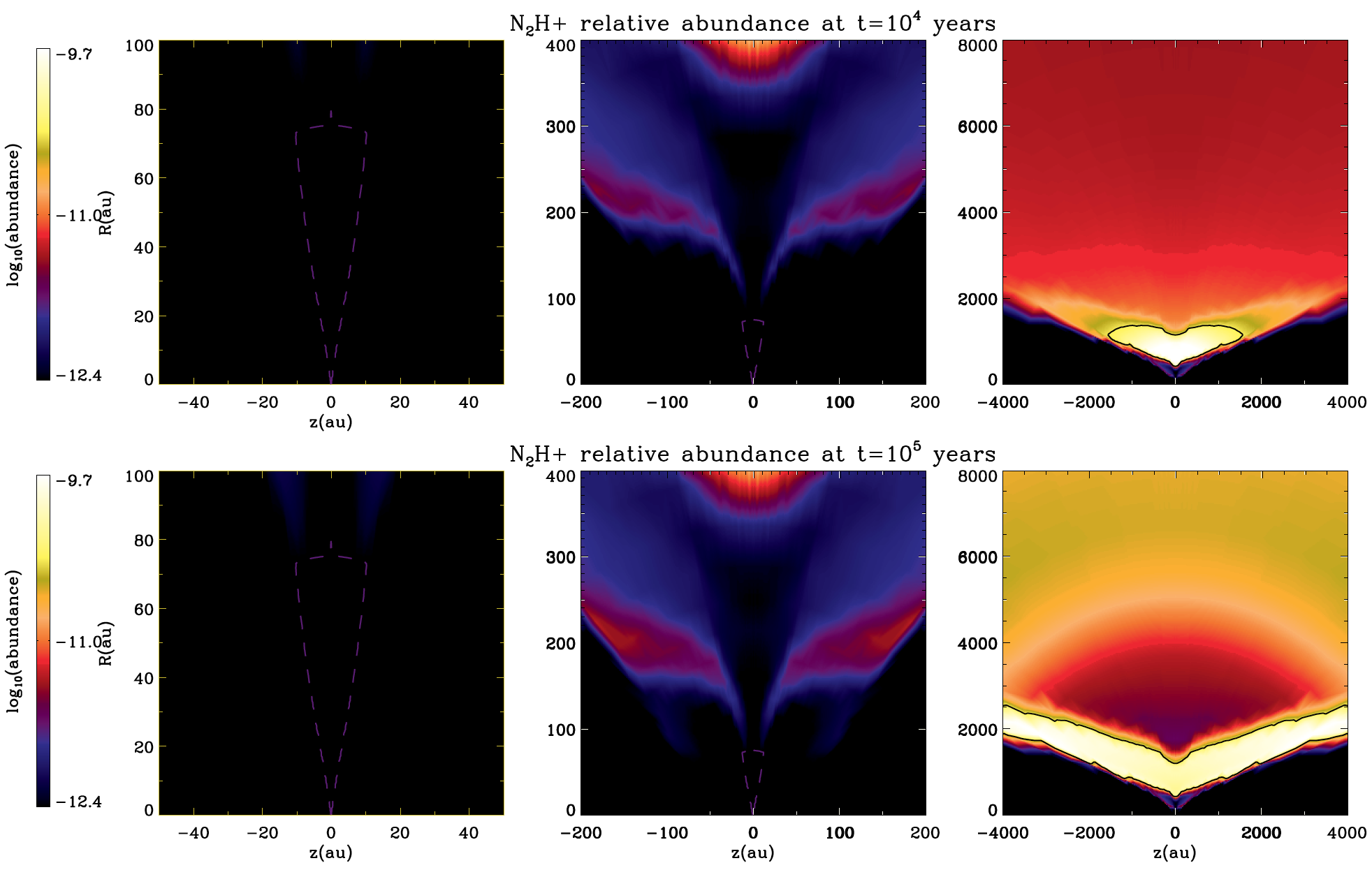}{0.8\textwidth}{(b)}
        }

\caption{Evolution of $^{12}$CO (Panel a) and of N$_{2}$H$^{+}$ (Panel b) gas phase abundance both for the TSC model. In Panel a, the black contour line shows the maximum $^{12}$CO abundance, divided by 10. In Panel b, the black contour line shows the maximum N$_{2}$H$^{+}$ abundance, also divided by 10. The purple dashed line in the first panel shows the disk. The maximum CO abundance is 7.22$\times$10$^{-5}$. The colorbar is in logarithmic scale.} \label{fig:n2hco}
\end{figure*}

\subsection{\texorpdfstring{C$^{18}$O abundance distribution}{}}

The time scale for freeze-out onto grains depends on both the temperature and density \citep{Lee_2004}, which vary throughout the cloud. The chemistry at the outer boundary of the cloud is determined by the cold temperature, low density, and exposure to the ISRF. Within the cloud envelope, at first the rapid increase of density inwards dominates, so that molecules freeze out onto grains in regions of low temperature and high density. Nearer the protostar the temperature rises above the desorption temperature, at which point the behavior changes, and molecules come off the grains and are released into the gas phase. The chemistry is also affected by the UV radiation field from the protostar. Dust extinction shields regions near the midplane from the protostellar UV field.

The dense gas distribution is often traced using isotopologues of CO, like C$^{18}$O, which are less likely to be optically thick than CO itself. For both time steps, $t=10^{4}$ years and $t=10^{5}$ years, high C$^{18}$O abundances ($\sim$10$^{-7}$ relative to hydrogen) are primarily coming from the disk and envelope at radii smaller than 400 au, where the temperature is $\geq 25$ K (see Fig.~\ref{fig:tempdens}) and the chemistry can be dominated by thermal desorption reactions and perhaps photodesorption reactions due to the capture of stellar radiation. Due to the lower abundance and optical depth of C$^{18}$O, the effect of photodissociation is visible by a drop of 10$\times$ in C$^{18}$O abundance already in the outer envelope ($\sim$8000 AU) at $t=10^{5}$ years. But in terms of spatial extension, the C$^{18}$O is similar to $^{12}$CO, as seen in Fig.\ref{fig:n2hco} panel a, but different for absolute abundance. 
We found no significant difference between UCM and TSC model for chemical abundances, and therefore limit the discussion to the TSC model abundances.

Our chemistry matches expectations in the envelope outside the edge of the disk, where we expect lower concentrations of C$^{18}$O ($\leq$10$^{-8}$) due to the high density area that is shielded from stellar radiation, and thus, cooler in this region. An abundance gap in the midplane beyond 300 au is present for both $t=10^{4}$ and $t=10^{5}$ years, and is consistent with the depletion expected in the region beyond the CO snowline ($T\leq$25~K) in the disk and envelope midplane. At $t=10^{5}$ years there is a steep drop in the C$^{18}$O abundance between 400 au and 4000 au due to the short freeze-out timescale compared with the outermost part of the cloud where the freeze-out timescale becomes longer \citep[i.e.,][]{caselli_1999}.

In the outermost region in the envelope ($r >$ 4000~au), the temperature drops to $T=10$ K; the light enhancement of C$^{18}$O abundance ($\sim$10$^{-8}$) in this region is influenced by photodesorption reactions such as CRs coming from outside the cloud. 

The chemistry results presented here for $^{12}$CO, C$^{18}$O, and also for N$_{2}$H$^{+}$ (next section \S\ref{sec:n2h+tracer}) represent a small subset of molecular abundances that are available for comparison with observations. Our RadChemT model includes a chemical network that provides abundance predictions for 292 chemical species; this number includes carbon and oxygen isotopologues and 90 grain surface abundances, a promising venue for further studies.

\subsection{\texorpdfstring{N$_{2}$H$^{+}$ as CO snowline tracer}{}}
\label{sec:n2h+tracer}

Snowlines, such as those for water and CO, are important because they can influence the efficiency of planet formation within the cold shielded regions of disks. Because CO line emission is optically thick thus making it difficult to observe the dense interior regions, recent studies by \citet{Aikawa_2015} and \citet{Hoff_2017} suggest that it is necessary to derive the location of the CO snowline from N$_{2}$H$^{+}$ observations. Observations have established that the N$_{2}$H$^{+}$ abundance is anti-correlated with CO abundance in systems ranging from starless cores (i.e, \citet{Tafalla_2004}), to class 0 protostars (i.e. \citet{Jorgensen_2004}), to protoplanetary disks (i.e. \cite{Qi_2013,2015ApJ...813..128Q, 2019ApJ...882..160Q,2019MNRAS.485.1843W}). 

Our chemistry results show a general predicted trend, that gas phase $^{12}$CO and N$_{2}$H$^{+}$ are anti-correlated in the envelope. In Figure~\ref{fig:n2hco} (Panel $a$, top middle and right) at $t=10^{4}$ years the $^{12}$CO is depleted from the gas phase and at $t=10^{5}$ years (Panel $a$, bottom middle and right) the $^{12}$CO depletion region increases outwards. The increase of N$_{2}$H$^{+}$ concentration, beyond 400 au, is consistent with the $^{12}$CO depletion to $t=10^{5}$ years (Figure~\ref{fig:n2hco} (Panel $b$, bottom middle and right). Inside 320 au, $^{12}$CO is enhanced since T$>$25 K in this region (see the first two top panels in Fig.\ref{fig:tempdens}) due to thermal and photodesorption reactions whereas N$_{2}$H$^{+}$ is absent. Therefore, the $^{12}$CO snowline is present and the best anticorrelation takes place at $T\sim25$ K in the midplane.

The predicted anti-correlation covers a larger region at $t=10^{5}$ years, which is the nominal age of L1527, and shows that the N$_{2}$H$^{+}$ concentration grows and extends from 400 au to $\sim$2000 au (close to the midplane). The N$_{2}$H$^{+}$ is present in the outer envelope at larger abundances at $t=10^{5}$ years compared to $t=10^{4}$ years.
Thus, our models show a general predicted trend, that gas phase $^{12}$CO and N$_{2}$H$^{+}$ are anti-correlated in the envelope. Although the CO depletion timescale differs somewhat from the N$_{2}$H$^{+}$ growth timescale, the N$_{2}$H$^{+}$ abundance increase should closely follow the slow depletion of $^{12}$CO since gas-phase timescales are much less than freeze-out timescales. The chemistry models are thus consistent with N$_{2}$H$^{+}$ being a good tracer for the $^{12}$CO snowline. Note that the abundances in the outflow cavity are excluded from consideration (see \S\ref{sec:chemicalmodel}). The jagged boundary between outflow cavity and envelope is due to grid sampling effects mentioned in \S\ref{sec:chemicalmodel}.

\subsection{Spectral line comparison: models versus observations}

In order to compare our models with observations, we generate synthetic model spectral line images of N$_{2}$H$^{+}$ for CARMA and compare its anti-correlation with $^{12}$CO CARMA data (\S\ref{sec:n2h+_carma}). In order to convey the validation of RadChemT, we generate a synthetic spectral line for C$^{18}$O, from which we extract synthetic P-V diagrams to compare with ALMA observations (\S\ref{sec:c18o_alma}). Each model spectral line data cube is based on the density, temperature, velocity grids, and chemical abundances analyzed from previous sections.

\subsubsection{\texorpdfstring{N$_{2}$H$^{+}$ and $^{12}$CO CARMA}{}}
\label{sec:n2h+_carma}

\begin{figure*}[ht!]
\centering
\includegraphics[scale=0.35]{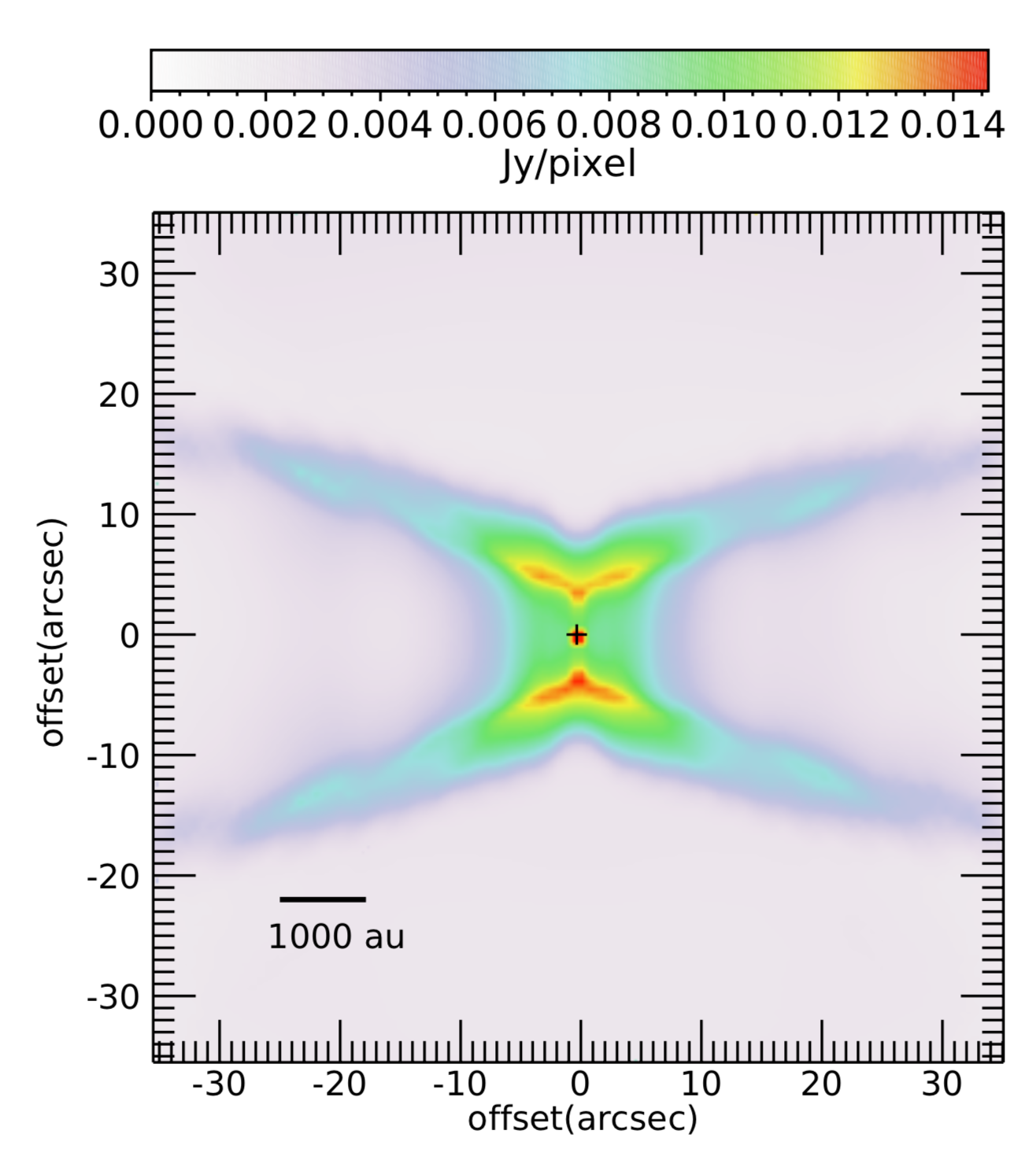}
\caption{The predicted model emission of N$_{2}$H$^{+}$ at $t=10^5$ years in units of Jy/pixel, having $0.67\arcsec$ pixel size. The + symbol at the center shows the protostar position. The disk diameter is only $\sim$1 pixel (smaller than the + symbol). Velocity channels covering the main hyperfine complex from 4.9 to 7.0~km~s$^{-1}$ contribute to the integrated intensity. The model was not convolved with the 9$\arcsec$ beam in order to more clearly show the spatial structure. Notice that the predicted N$_{2}$H$^{+}$ emission peaks north and south of the protostar position.  \label{fig:n2h+model}}
\end{figure*}

\begin{figure*}[ht!]
\gridline{\fig{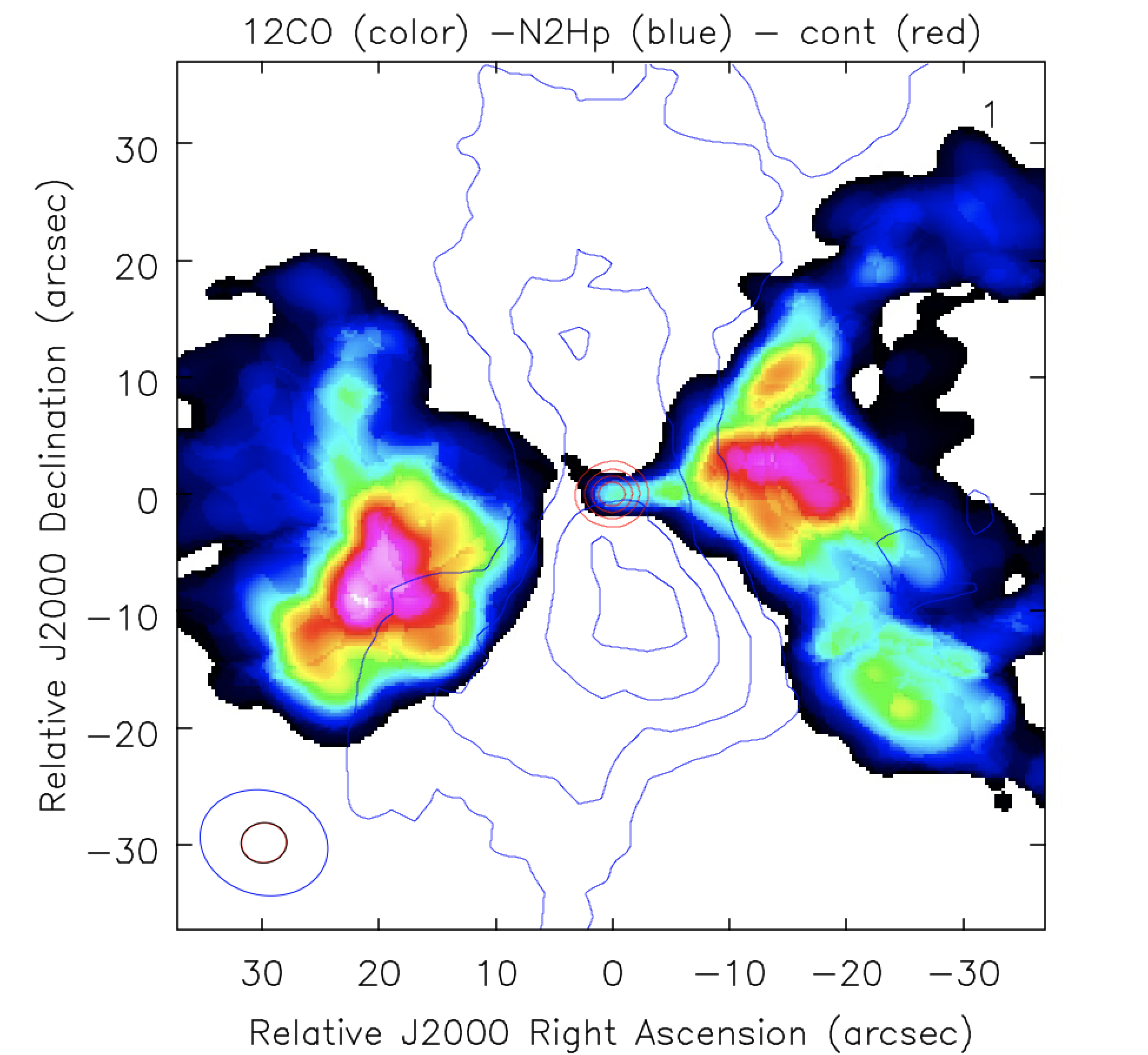}{0.45\textwidth}{(a)}
        %\centering
          \fig{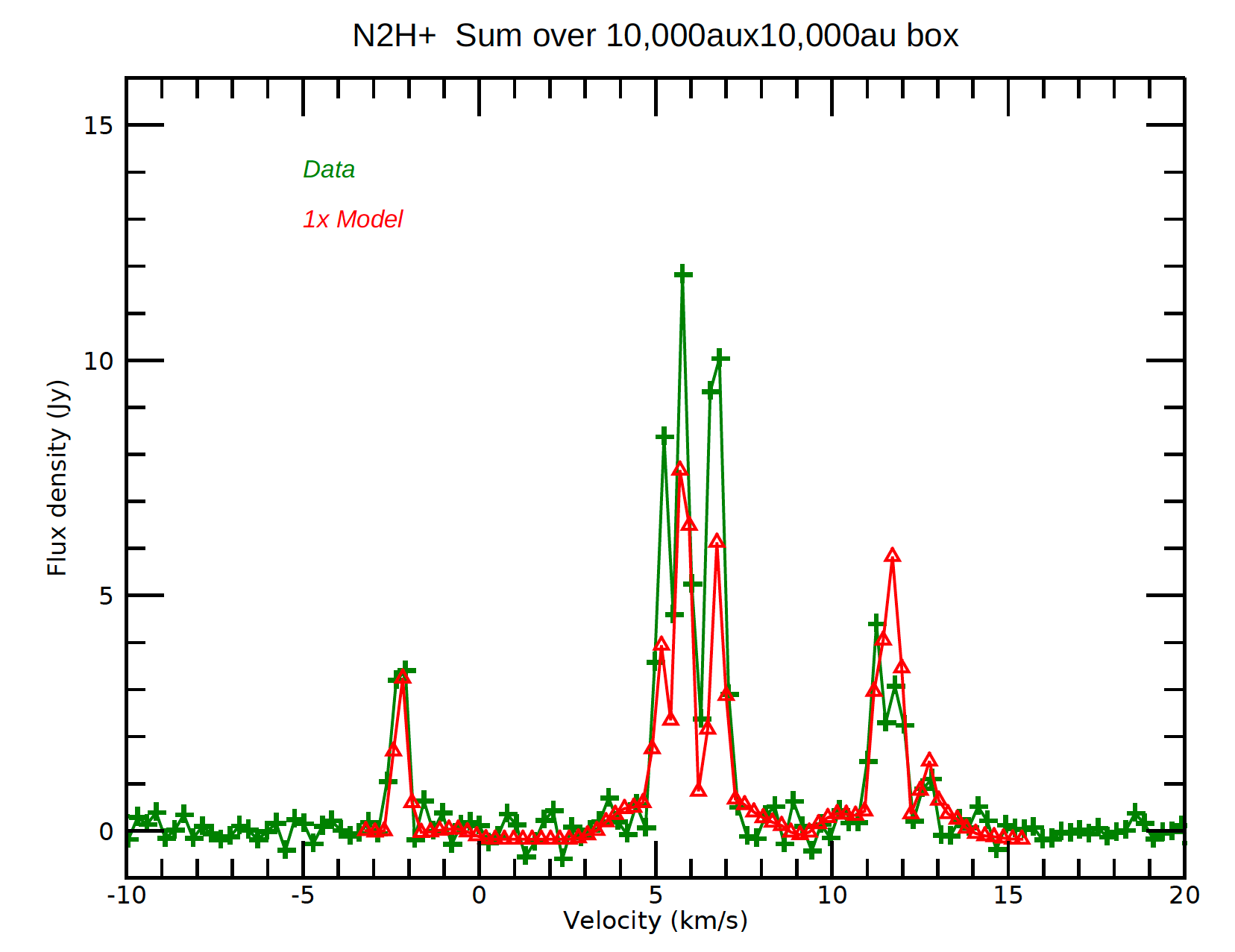}{0.5\textwidth}{(b)}
         }
\caption{$Left$: L1527 composite from CARMA data. Image size is $71.4\arcsec$ ($10000$ au). The color image shows integrated $^{12}$CO emission, which is seen to trace the outflow in the east-west direction. The red contour lines are dust continuum emission centered on the protostar position. The blue contour lines show the dense gas distribution in N$_{2}$H$^{+}$ emission, which extends north-south. Notice that the N$_{2}$H$^{+}$ emission peaks at roughly $10\arcsec$ north and south of the protostar position. The CO and N$_{2}$H$^{+}$ beams are drawn in the lower left corner of the image. $Right$: shows the N$_{2}$H$^{+}$ (1-0) spectrum, including hyperfine components, constructed using a 10,000 au size box centered on the protostar position. Green plus symbols are CARMA data, and red line shows the RadChemT model prediction.  \label{fig:spitzercarma}}
\end{figure*}

Figure~\ref{fig:n2h+model} presents the RadChemT model in the form of an integrated intensity map of N$_{2}$H$^{+}$ at $t=10^5$ years. The best N$_{2}$H$^{+}$ and $^{12}$CO anticorrelation takes place at $t=10^5$ years, where the greater extension of the N$_{2}$H$^{+}$ abundance (see Panel $a$ in Fig.~\ref{fig:n2hco}) is more consistent with the CARMA data than at $t=10^4$ years. In Figure~\ref{fig:n2h+model} we predict the spatial extension of the N$_{2}$H$^{+}$ intensity. By not convolving the modeled emission of Figure~\ref{fig:n2h+model} with the beam, we are able to see in more detail the spatial structure of the N$_{2}$H$^{+}$ emission in the envelope, otherwise, the X-shape emission would disappear looking more like a vertical bar, very similar to how it looks in Figure~\ref{fig:spitzercarma} Panel (a). As previously discussed in \S\ref{sec:n2h+tracer}, the predicted N$_{2}$H$^{+}$ emission traces cold dense gas in the envelope between $400$ au and $\sim 3000$ au, extending north-south along the midplane. The predicted emission is seen to peak north and south of the protostar position, with less emission at $\leq400$ au, or about $ 3\arcsec$ in radius, where Figure~\ref{fig:n2hco} predicts that N$_{2}$H$^{+}$ should be absent. Although less visually prominent, there is also extended N$_{2}$H$^{+}$ that corresponds to the outer envelope, that is faintly seen in Figure~\ref{fig:n2h+model} at the level of $\sim 0.002~Jy~pixel^{-1}$, extending across the simulated image.

The N$_{2}$H$^{+}$ spectrum is constructed by integrating over the entire $10,000$ au image, and which moreover includes and confirms the hyperfine components seen in Figure~\ref{fig:spitzercarma} Panel (b). The red solid line (model) and green plus symbols (data) show good correspondence at $t=10^{5}$ years. Note that the extended (faint) emission  contributes significantly to the predicted spectrum. For the two emission peaks from the inner envelope (two red peaks in Fig.\ref{fig:n2h+model}), there is a discrepancy between the spatial extent of our prediction (Fig.~\ref{fig:n2hco} and Fig.~\ref{fig:n2h+model}) and the observation (Fig.\ref{fig:spitzercarma}) that we conjecture might be improved by increasing the collapse age of the system in terms of changing the dynamics so that $R_{col}$ leads to a larger collapsing region. This is a good motivation to follow a time-dependent evolutionary parcel and, thus, confirm the spatial extension. Overall, the strength of the observed N$_{2}$H$^{+}$ emission is consistent with the model prediction as seen in the spectrum.

Figure~\ref{fig:spitzercarma} also presents the integrated intensity maps of $^{12}$CO (1-0) from CARMA that cover the same 71.4\arcsec\ ($10,000$ au) field of view and that shows out-flowing CO gas that extends east-west, perpendicular to the N$_{2}$H$^{+}$ emission. The dust continuum emission (red contours) is centered on the protostar position. In $^{12}$CO a narrow ``jet" extends east-west from the protostar, merging into a wide-angle outflow on both sides of the protostar. The narrow ``jet" is also confirmed by ALMA (Fig.\ref{fig:coalma}), therefore, supporting the inclusion of an inner outflow jet in our model. Fig.\ref{fig:cochannels} in Appendix C contains the $^{12}$CO velocity channel maps, that further show that the emission arises in an ($\pm$ 3~km~s$^{-1}$) outflow shell. None of the emission extends north-south, since the high optical depth of $^{12}$CO blocks any meaningful view of the low-velocity envelope.

We conclude that the N$_{2}$H$^{+}$ emission shows evidence for the predicted anti-correlation with $^{12}$CO. The N$_{2}$H$^{+}$ appears to be missing within $400$ au of the protostar, just where full strength $^{12}$CO emission is predicted. One difference is that our predicted emission is less extended compared with the data (Fig.~\ref{fig:n2h+model}), that extends out to $\sim$30$\arcsec$ ($\sim$4200~au). However, the data agree in showing N$_{2}$H$^{+}$ enhancement at $\pm 18\arcsec \sim 2500$ au from the protostar, where the model predicts there is severe depletion for $^{12}$CO (see Fig.\ref{fig:n2hco}). A factor that might help to further improve the comparison is to increase $R_{col}$ and follow the evolutionary process of the chemical parcel spatially.

\subsubsection{\texorpdfstring{C$^{18}$O ALMA}{}}
\label{sec:c18o_alma}

We focus on C$^{18}$O (2-1) observations from ALMA in order to investigate the kinematics of the dense gas distribution in the disk and inner envelope in a spectral line tracer that is (nearly) optically thin. The orientation of the rotating and infalling material is extended from north-south (as seen in Fig.~\ref{fig:spitzerandschematic}) and C$^{18}$O probes dense gas near the protostar. The edge-on inclination of L1527 means that a north-south cut along the midplane will minimize the contamination of the dense gas emission by the outflow. However, the central $0.8\arcsec$ beam, meaning $\pm0.4\arcsec$ centered on the protostar, can still contain some outflow emission. The $^{12}$CO emission (Fig.\ref{fig:coalma}) for the same size region is optically thick and traces lower density gas extending east-west in the outflow shell.

To compare with the ALMA data we make a synthetic model spectral line cube that matches the ALMA data file. Namely, we generate velocity channel maps for C$^{18}$O (2-1) with a velocity spacing of 0.167~km~s$^{-1}$ and $0.17\arcsec$ pixel size. We convolve the model with the $0.8\arcsec$ effective beam, and apply the $28\arcsec$ primary beam, as described in \S\ref{sec:modelvisualization}. Note that RadChemT includes a basic description of the outflow velocity prescription of 3~km~s$^{-1}$ in the model, since RadChemT self-consistently computes both the abundances of $^{12}$CO and C$^{18}$O as well as the radiative transfer of the two species (LTE is assumed).

\begin{figure*}
\gridline{\fig{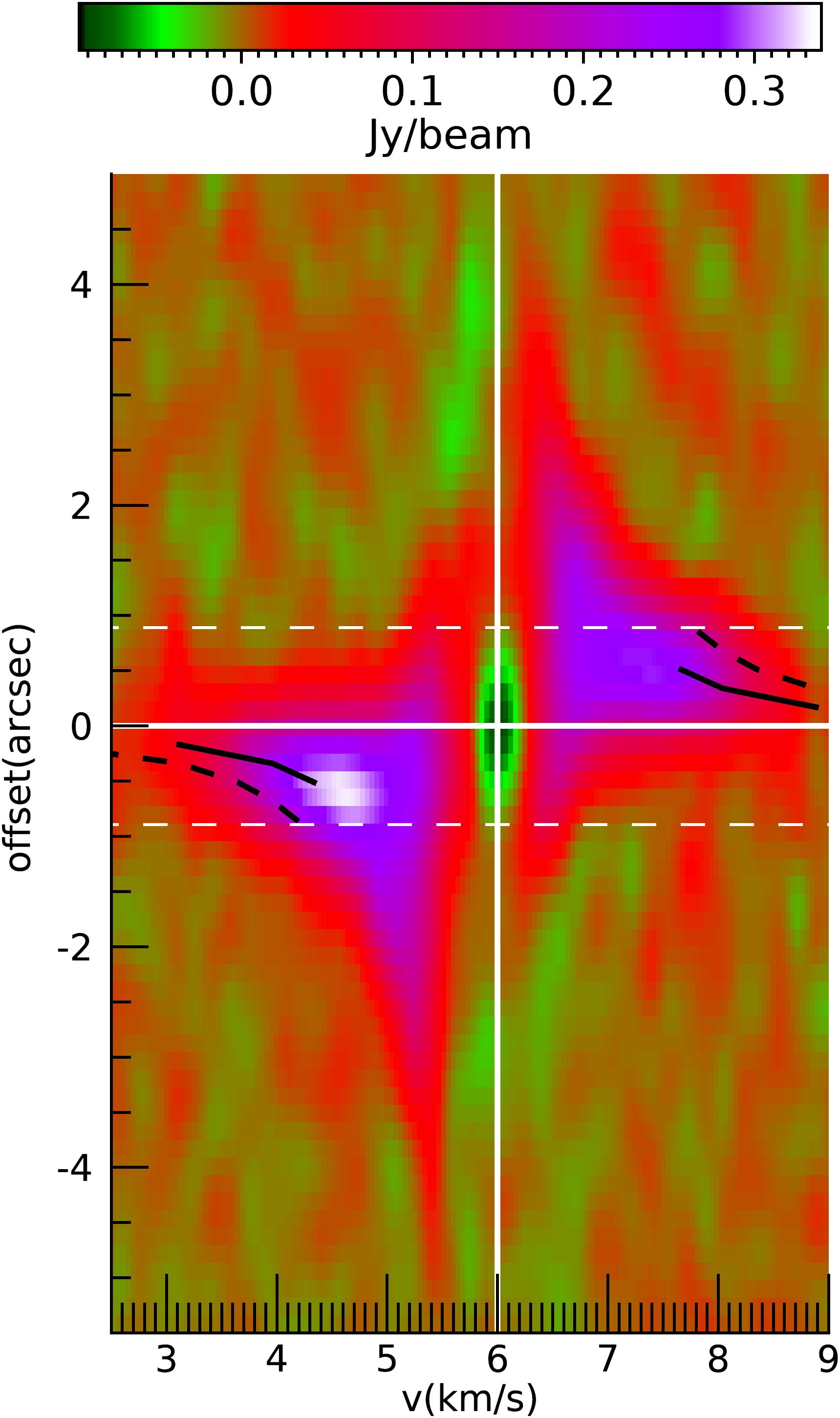}{0.3\textwidth}{(a)} 
          \fig{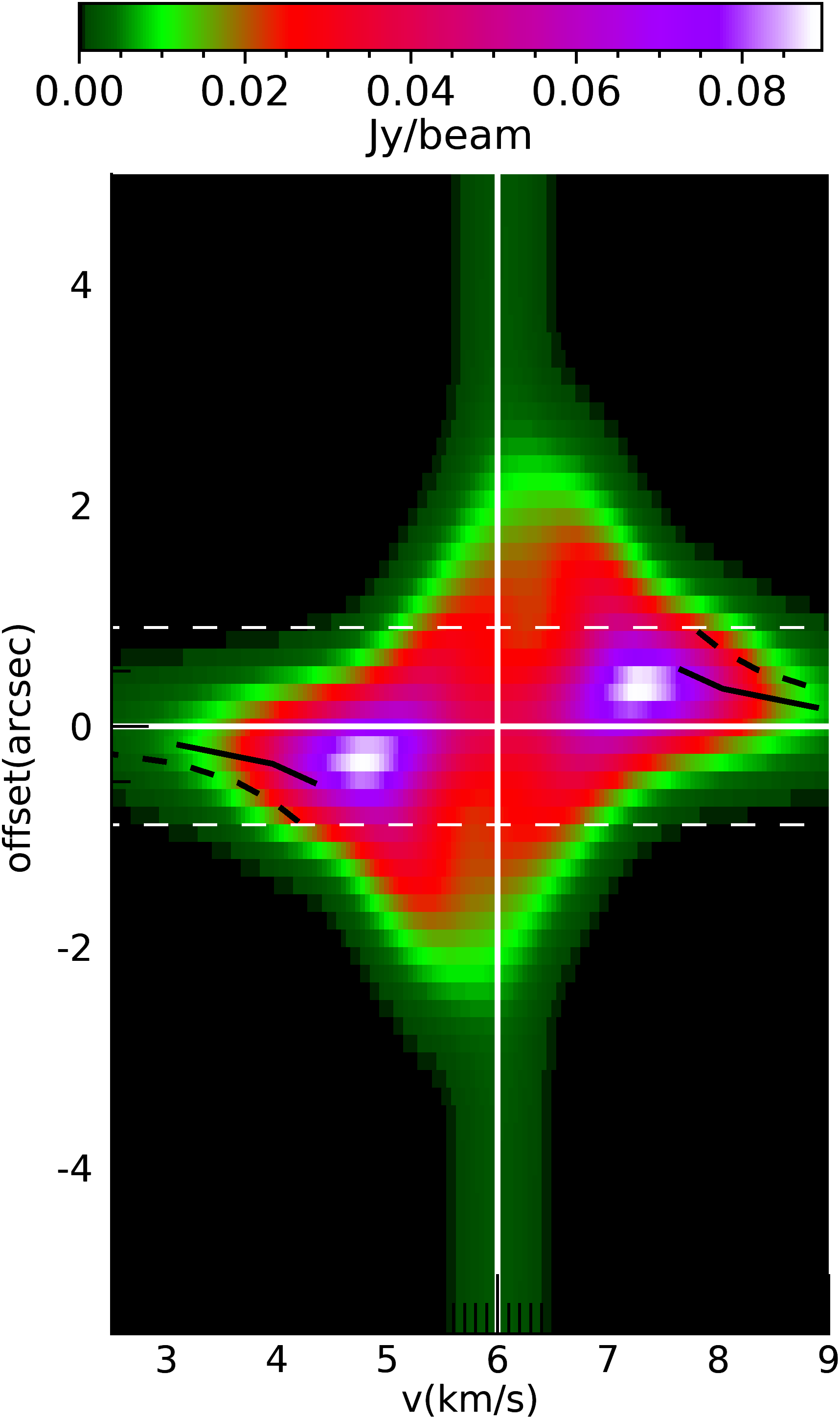}{0.3\textwidth}{(b)}
         \fig{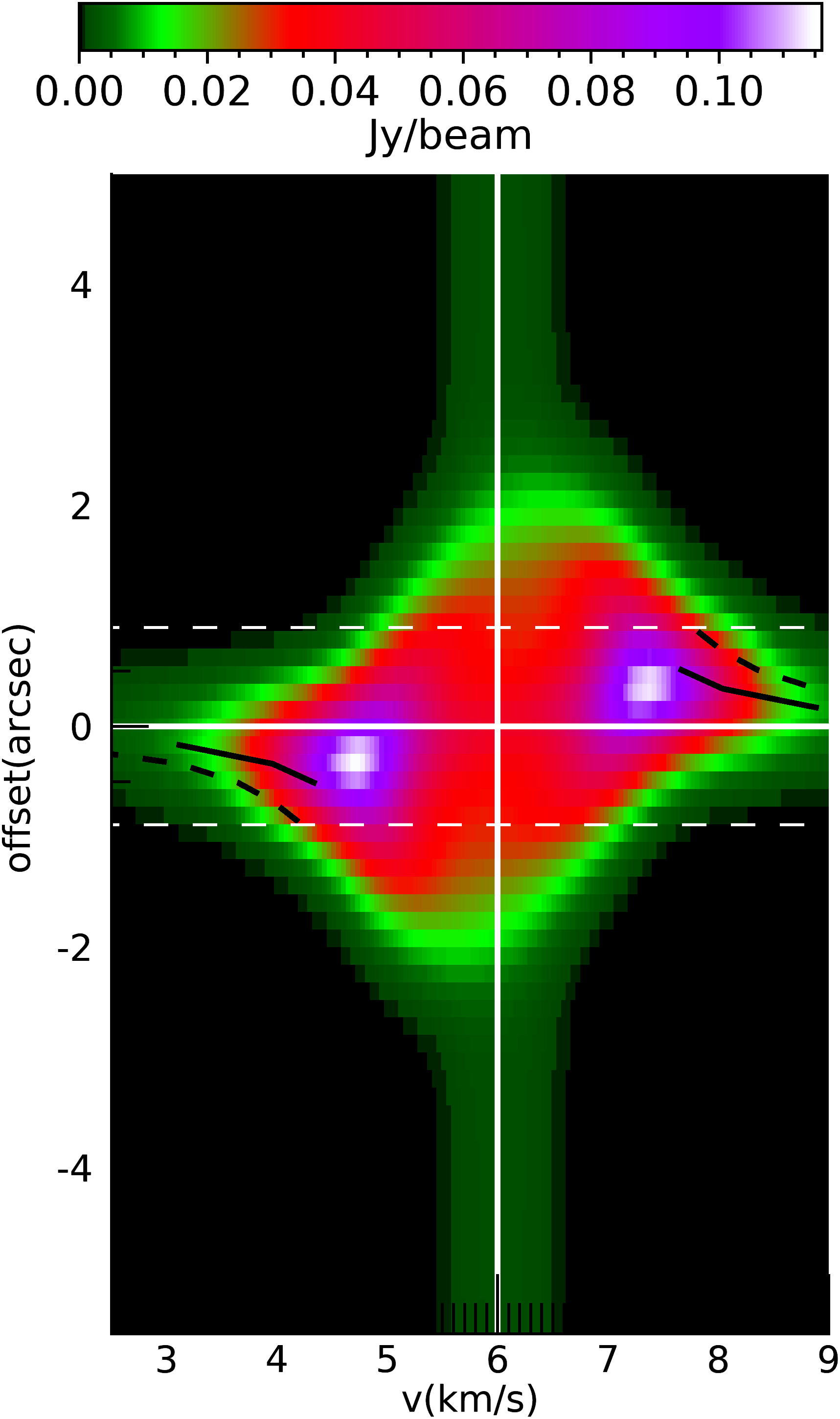}{0.3\textwidth}{(c)}
         }
\gridline{\fig{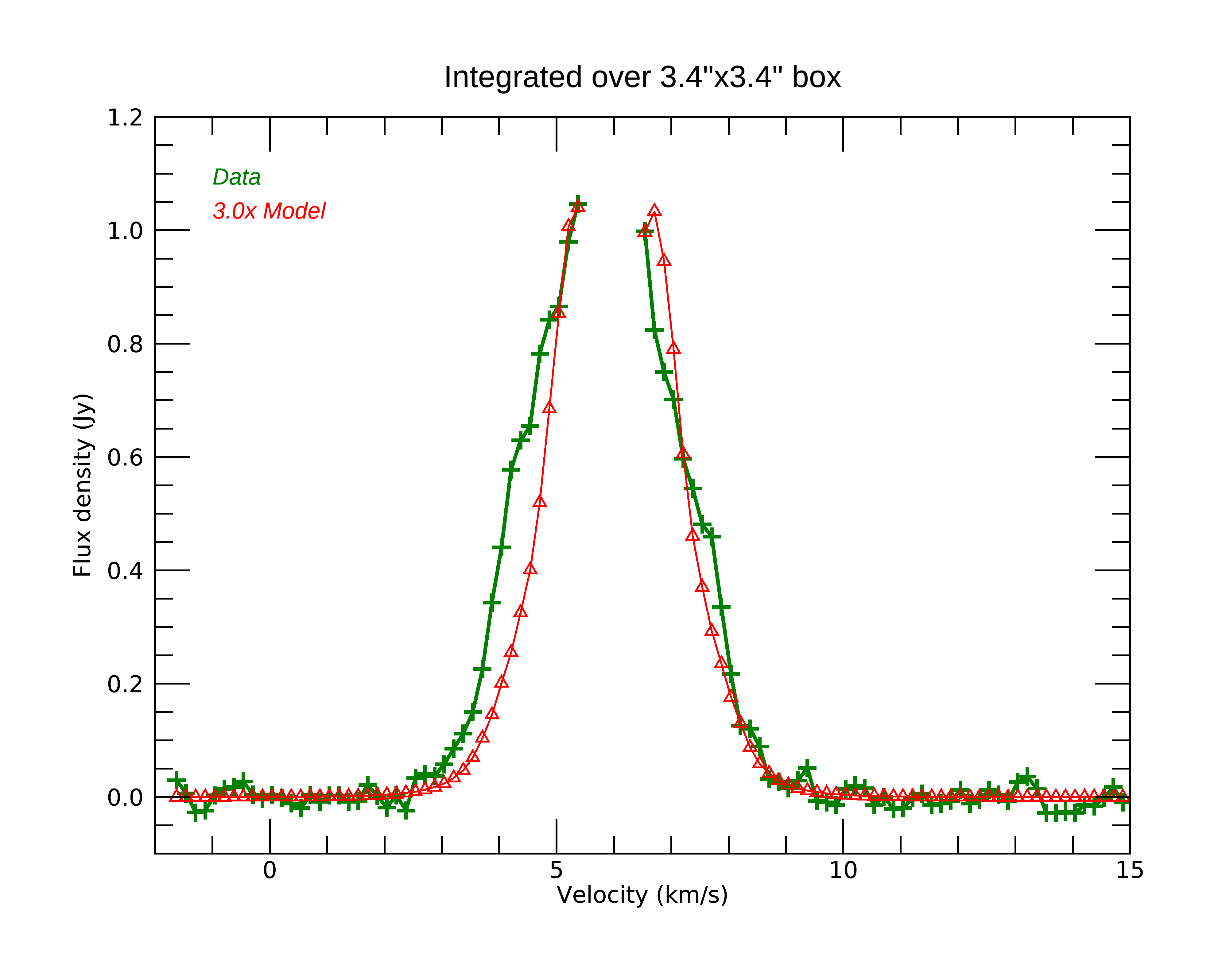}{0.5\textwidth}{(d)}
} 
\caption{Position-Velocity (PV) diagrams that show the velocity of C$^{18}$O versus position offset along the north-south direction. Each horizontal row corresponds to a spectrum at the indicated spatial offset. Panel (a) shows the ALMA C$^{18}$O observational data. The P-V diagram obtained from the RadChemT chemical model is shown in panel (b) for the TSC collapse model and panel (c) for the UCM collapse model.  The white solid line in each panel shows the system radial velocity (vertical) and central protostar position (horizontal). Dashed white lines represent $\pm$125au, the maximum disk size considered. Keplerian rotation curves are included for reference. The solid black line represents the fiducial model with $M_{*}$=0.22 $M_{\odot}$ and $R_{disk}$=75~au. The dashed black line is $M_{*}$=0.45 $M_{\odot}$ and $R_{disk}$=125~au. Panel (b) and Panel (c) shows different color scale bar due to the higher density of UCM in the inner envelope.
Panel (d) shows the C$^{18}$O (2-1) spectrum, integrated over a 3.4 $\arcsec \times$ 3.4 $\arcsec$ box centered on the protostar. Green plus symbols are ALMA data; red triangles show the RadChemT TSC collapse model, multiplied by a 3.0$\times$ scaling factor. 
\label{fig:pv}}
\end{figure*}

The edge-on inclination of L1527 means that a PV diagram is well-suited for viewing the kinematics of rotation and infall in the central envelope and disk. Figure~\ref{fig:pv} presents PV diagrams for C$^{18}$O (2-1), where panel $a$ shows the ALMA data that is also presented in \citet{Ohashi_2014}. These data are sampled in real space to clearly see the cloud features in more detail. Panel $b$ shows the simulated L1527 as modeled using the TSC model and, panel $c$ shows the simulated L1527 as modeled using the UCM model. Each horizontal row in Fig.~\ref{fig:pv} corresponds to a spectrum that is spatially offset from the protostar, in the north-south direction. The horizontal solid white line marks the position of the protostar, which is located at $0\arcsec$ offset. The effective width of the position slice is the $0.8\arcsec$ beam (=112~au @140pc) and the units are $Jy~beam^{-1}$. The vertical solid white line shows the adopted 6.0~km~s$^{-1}$ Doppler radial velocity of L1527. 

The data (panel $a$) show that C$^{18}$O is self-absorbed at the protostar position, where the solid white lines meet at the center of the PV diagram. This indicates that the C$^{18}$O is optically thick at line center, consistent with the finding of \citet{van_t_Hoff_2018}. However, our model does not currently reproduce the self-absorption feature, seen to occur within $\pm$0.5 km~s$^{-1}$ from line center (6.0~km~s$^{-1}$). In a later paragraph we further discuss the self-absorption. The data (panel $a$) also show an artifact that is common to interferometers. The cloud emission from L1527 is spatially extended and therefore resolved-out by the interferometer at the 6.0~km~s$^{-1}$ cloud velocity, resulting in no/little emission from the cloud near the vertical line. However, the RadChemT images retain the low-velocity cloud emission at 6.0~km~s$^{-1}$, and do not mimic this artifact of the data.

At large spatial offsets ($>4\arcsec$) the spectral line has a narrow width that approximates the thermal sound speed of the cloud. The range of velocities $v$ in each spectrum  (i.e. horizontal row) increases towards the protostar (i.e. smaller R and smaller position offset), as is expected for gravitational motion where $v\sim \sqrt{GM/R}$. The emission occurs mainly in the lower left and upper right quadrants, which is the expected signature of rotational motion in a disk. Emission that occurs in the ``forbidden'' quadrants (upper left and lower right) is not from the rotating disk, but instead is a signature of the rotating and infalling envelope (i.e. \citet{ho_keto_2007}).

Comparison of the PV diagrams in Fig.~\ref{fig:pv} shows an overall correspondence of the data with the RadChemT models that is encouraging. Visually, the TSC model (panel $b$) is a better fit than the UCM model (panel c). One difference between data and model is that the peak brightness values are symmetric in the models but not symmetric in the data.  The data (panel a) show stronger blue shifted peak emission (lower left quadrant) than red shifted (upper right quadrant), which is a signature of optically thick emission. To improve the model fit, this suggests that the model envelope density (or abundance) should be increased over the fiducial value.

Recent analyses of L1527 in the literature using different datasets find $M_* = (0.19,0.45,0.45~M_\sun$) and $R_{disk}$= (125,75,125~au), respectively \citep{tobin_2012,Aso_2017,van_t_Hoff_2018}. In our fiducial model we adopted the \citet{tobin_2012} values of $M_* = 0.22~M_{\sun}$ and $R_{disk}= 75$~au as a starting point to test the capabilites of RadChemT on reproducing the chemical abundances of L1527. Our fiducial numbers fall at the lower end of recent determinations. The analysis by \citet{van_t_Hoff_2018} is the most similar to our modeling, although it assumes but does not compute the astrochemical abundances, and moreover restricts attention to the inner $\sim 1\arcsec$(140~au).

In all three PV diagrams, the dashed white line represents the maximum size of the disk. Everything outside of the dashed white lines, at $> 1\arcsec$ position offset, is emission coming purely from the envelope. The emission coming from inside these lines is therefore due to a combination of envelope and disk emission. Keplerian rotation curves are presented as a guide for the eye, and represent the {\it{maximum}} velocity expected from a rotating Keplerian disk. The solid black line represents the fiducial model with $M_{*}$=0.22 $M_{\odot}$ and $R_{disk}$=75~au. The dashed black line is $M_{*}$=0.45 $M_{\odot}$ and $R_{disk}$=125~au, a model that is considered in \citet{van_t_Hoff_2018}. Comparison of the model PV diagram with the data in Fig.~\ref{fig:pv} suggests the higher protostar mass would be preferred over the fiducial value. However, detailed model fitting of the disk dynamics lies outside the scope of the current work.

The strength of the observed C$^{18}$O(2-1) emission is about a factor of 3.0 higher than the model prediction for the TSC model. Figure~\ref{fig:pv} (panel $d$) 
shows the C$^{18}$O spectrum that is constructed by integrating over a $3.4\arcsec \times 3.4 \arcsec$ box (475~au $\times$ 475~au). The red triangles (model) and green plus symbols (data) show reasonable correspondence in terms of profile shape. The overall emission of the UCM model is about (50\%) brighter; this difference is understandable as due to the higher density of UCM in the {\it inner} envelope for our choice of physical parameters; an approximate estimate of the density ratio expected between UCM and TSC is {5/3}, and is simply obtained from the ratio of $\dot{M}_{env}$ that are given in Table~\ref{tab:Physical_Parameters}.

The fact of not reproducing the self-absorption at the protostar position in our modeled C$^{18}$O spectrum does not preclude the ability of the models to explain what is happening at the center of the spectral line. The observed self-absorption is consistent with a high density region, grains with millimeter size or greater, in the disk that suggests a promising avenue for future studies to increase the density between the envelope and disk.  Considering a different type of dust opacity that is more suitable for the inner parts is also encouraging. On the model processing side, performing the spectral tracing using non-LTE may also lead to improvement. In summary, the RadChemT model was not tweaked to match the C$^{18}$O emission, so the initial match between model and data to within the factor of 3.0 is encouraging. 

From Figure~\ref{fig:pv} we see that our prediction of C$^{18}$O using the TSC model is better at representing the actual envelope structure of L1527 when compared with C$^{18}$O ALMA data. Due to the fact that UCM neglects pressure effects in the outer layers of the envelope, we see that at spatial offsets $>1\arcsec$ (150~au), the UCM envelope shows too much gas at higher velocity, resulting in a rectangular rather than bowtie shape around the green perimeter showing the fainter emission. In general, we conclude that the predicted C$^{18}$O(2-1) emission from the RadChemT model reproduces the main features of the PV diagram (Fig.~\ref{fig:pv}) for both the envelope and disk emission. Moreover, these initial results suggest that RadChemT can be used as a tool to investigate the protostar dynamical mass, the disk radius, and the unknown dynamics of the outer disk. Future improvements to RadChemT to specifically model L1527 better could include: 1)increasing the density and opacity profile in the disk and performing non-LTE spectral line radiative transfer (i.e., \cite{Evans_1999}),  2)changing the physical conditions as a function of time to follow collapse motions, and 3)including shock physics and adding sulfur chemistry to study the suggested enhancement of SO at the disk-envelope interface.  

\section{Summary and conclusions} \label{sec:conclusions}

RadChemT is a method for modeling embedded protostars to compare with both, continuum and molecular line observations. The method combines a two-dimensional, varying both with distance from the star and angle from the rotation axis, axisymmetric cloud collapse solution with MCRT and the solution of an astrochemical reaction network. The resulting gas phase abundances are transformed via LTE radiative transfer into simulated Position-Velocity cubes to compare with spectral line observational data. This pilot study with RadChemT uses a model of the central star and surrounding gas density distribution obtained by \citet{Tobin_2008, Tobin_2010, tobin_2012} for the protostar L1527. In the current implementation, a protostar of a given age has time-steady density and temperature distributions, while the chemical abundance calculation is time-dependent. There are pronounced spatial variations in the abundances. Abundances are enhanced in the outflow shell and decreased in the cold regions near the midplane, varying by more than a factor of $10^3$ in the case of CO. In order to validate RadChemT, we generate PV diagrams for the inner 1,000~au ($14.2\arcsec$), and compare with ALMA C$^{18}$O observations. We also present CARMA data for $^{12}$CO and N$_{2}$H$^{+}$ on a larger 10,000~au scale, and compare with our predicted abundances for L1527. We report our highlights as follows:

\begin{enumerate}
    
\item The TSC and UCM collapse models give comparable fits to the SEDs, both for aperture size of 10,000~au and 1,000~au. Similarly, there is little qualitative difference for the predicted molecular abundances. However, the TSC model better corresponds to the observed PV diagrams. In the case of UCM, which neglects pressure forces, the envelope shows too much gas at higher velocity, particularly at spatial offsets greater than 150~au.

\item The ALMA C$^{18}$O (2-1) spectrum is about 3.0 times brighter than our C$^{18}$O prediction. This is reasonable agreement given that the astrochemical computation has not been ``tuned'' to improve the fit. However, increasing the density in the envelope and the opacity profile in the disk could be fruitful, since the C$^{18}$O abundance is sensitive to them. The dynamics of the C$^{18}$O gas imply that the protostar mass and disk radius are somewhat larger than the fiducial values of 0.22 $M_{\sun}$ and 75~au, respectively. 
 
\item The CARMA $^{12}$CO (1-0) data confirms that there is strong emission with the morphology of an outflow shell. The ALMA $^{12}$CO (2-1) data definitively establish that a narrow jet-like structure connects the two outflow lobes inside 75~au. For the physical model we therefore include a swept-up outflow shell with a constant outward velocity of 3 km s$^{-1}$ (Fig.\ref{fig:cochannels}) as a proof of concept. The chemistry implies that the $^{12}$CO abundances are low in the inner envelope from 400~au $<$ R$_{env}$ $<$ 2,000~au at $t$=10${^5}$ yrs, indicative of freeze out onto grains. 
 
\item In the CARMA N$_{2}$H$^{+}$ (1-0) data, emission is elongated north-south, with the peak emission offset $\sim 10 \arcsec$ from the central star. In our chemical model, N$_{2}$H$^{+}$ is also offset, by about $\sim 10 \arcsec$ north of the central star. This is indirect but strong evidence of significant $^{12}$CO freeeze out in the same region. As found in many previous studies, the chemistry implies that N$_{2}$H$^{+}$ is anti-correlated with CO abundance. In the case L1527, RadChemT predicts that N$_{2}$H$^{+}$ is enhanced in the envelope over 500 au $<$ R$_{env}$ $<$ 2,000 au at $t$=10$^{5}$ years.
 
\end{enumerate}

\facilities{ALMA, CARMA, HERSCHEL, IRSA, Spitzer, TIFKAM, IRAC, MIPS, SCUBA, JCMT/UKT, NMA}

\acknowledgements
I am very thankful to Dr. Susan Terebey, who helped tirelessly on this project. Very special thanks to all my co-authors as well. To Dr. Hengchun Ye and Dr. Krishna Foster for sponsoring through the NASA-DIRECT STEM program (Grant: NNX15AQ06A) and through the MORE RISE-to-PhD program (Grant: 2R25GM061331-18). This work was also carried out in part at the Jet Propulsion Laboratory, under contract with NASA and with the support of Exoplanets Research Program grain 17-XRP17$\_$2-0081. This project received support from the European Research Council (ERC) under the European Union$^{'}$s Horizon 2020 research and innovation programme (grant agreement 757957).  
To Dr. Andrea Isella (RICE) for providing the CARMA observational data. This paper makes use of the following ALMA data: ADS/JAO.ALMA\#2011.0.00210.S. 

ALMA is a partnership of ESO (representing its member states), NSF (USA) and NINS (Japan), together with NRC (Canada), MOST and ASIAA (Taiwan), and KASI (Republic of Korea), in cooperation with the Republic of Chile.

The Joint ALMA Observatory is operated by ESO, AUI/NRAO and NAOJ. The National Radio Astronomy Observatory is a facility of the National Science Foundation operated under cooperative agreement by Associated Universities, Inc.

\appendix
\section{\texorpdfstring{$^{12}$CO (2-1) ALMA channel map}{}}
\begin{figure}[ht!]
\centering 
\includegraphics[scale=0.8]{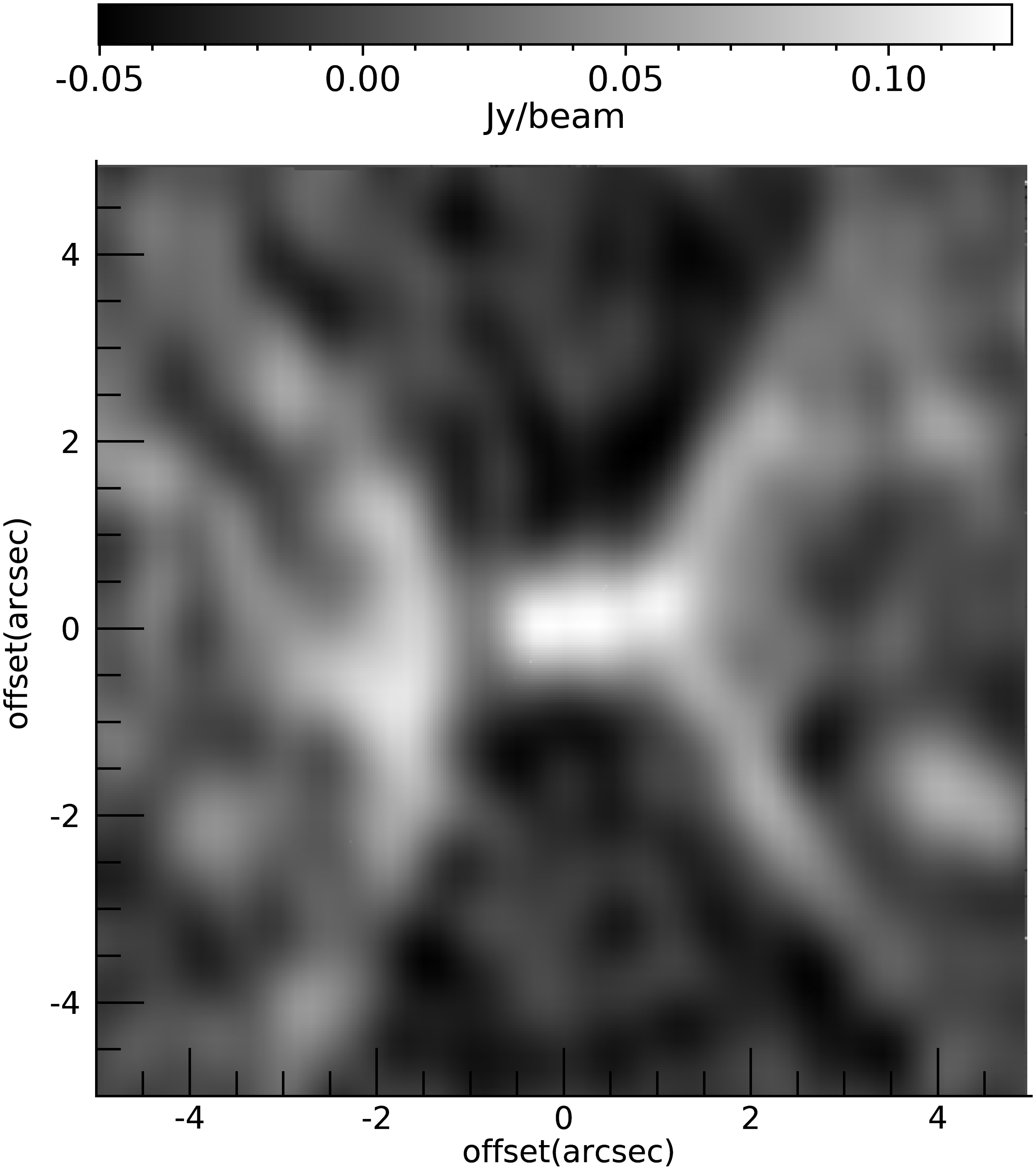} 
\caption{Snapshot of a $^{12}$CO(2-1) ALMA channel map at 9.9 km~s$^{-1}$ that shows the inner region of the outflow. The $^{12}$CO emission is shaded in white and shows two outflow shells lying in the east-west direction that are connected by a narrow jet. This orientation matches with Fig.\ref{fig:spitzerandschematic}, Panel (b). The data have a spatial resolution of 0.8\arcsec and the star position is at the center of the 10\arcsec (1400~au) image.
\label{fig:coalma}}
\end{figure}

\section{\texorpdfstring{CARMA $^{12}$CO channel maps}{}}
\begin{figure*}
\gridline{\fig{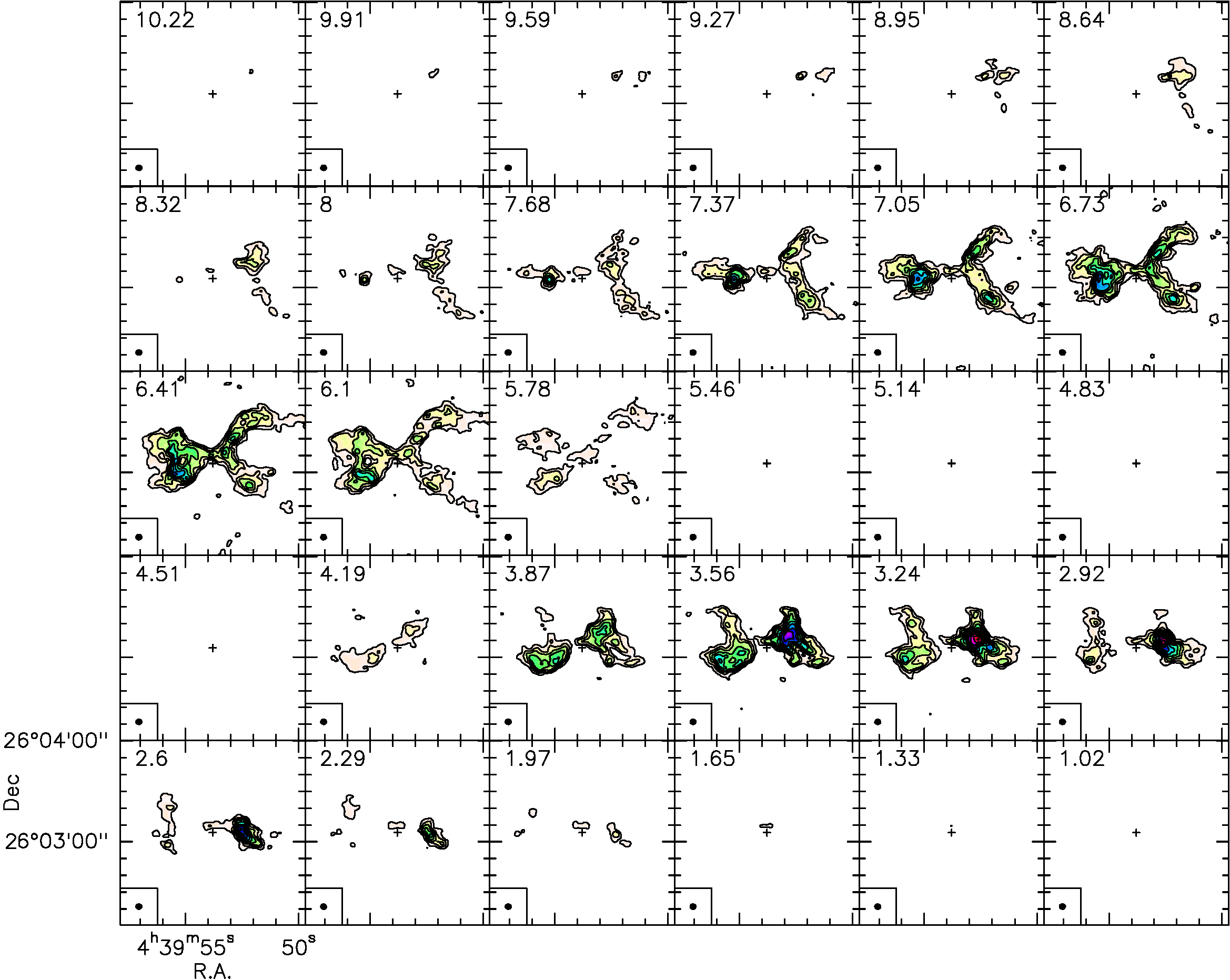}{1.0\textwidth}{}}
\caption{$^{12}$CO CARMA channel maps. Contours are spaced by 3$\sigma$ = 0.187 Jy/beam. Lower left inset shows the FWHM beam size 10.97"$\times$8.73". Each panel is 110"$\times$110". The radial velocity of L1527 is $6~km ~s^{-1}$; several blank velocity channels near $5~km ~s^{-1}$ suggest absorption due to a foreground cloud. In $^{12}$CO a narrow ``jet" extends east-west from the protostar, merging into a wide-angle outflow on both sides of the protostar. The $^{12}$CO velocity channel maps show that the emission arises in an ($\pm$ 3~km ~s$^{-1}$) outflow shell.
\label{fig:cochannels}}
\end{figure*}

\section{Initial abundances}

\begin{table}[ht]
\tablenum{4}
\begin{center}
\caption{Initial abundances. \label{tab:initabu}}
\begin{tabular}{cc}
\hline \hline
\textbf{Species} & $X$ \\
\hline  
H & 1.00$\times$10$^{-2}$\\
H$_{2}$ & 0.495 \\
He & 0.140 \\
C$^{+}$ & 7.22$\times$10$^{-5}$ \\
$^{13}$C & 8.11$\times$10$^{-7}$ \\
N & 2.14$\times$10$^{-5}$ \\
O & 1.75$\times$10$^{-4}$ \\
$^{17}$O & 8.80$\times$10$^{-8}$\\
$^{18}$O & 3.51$\times$10$^{-7}$ \\
Si$^{+}$ & 2.00$\times$10$^{-8}$ \\
\hline
\multicolumn{2}{@{}l@{}}{\scriptsize *$X$ is the fractional abundance relative to hydrogen, $n_{X}/(n_{H}+2n_{H{_2}})$.} \\
\end{tabular}
\end{center}
\end{table}

\section{ALMA observational parameters}
\section{Photometry}

\begin{center}
\begin{deluxetable*}{ccc}
\tablenum{5}
\caption{ALMA observational parameters.\label{tab:alma_obs_parameters}}
\tablehead{
\colhead{} & \colhead{C$^{18}$O(2-1) }\\
\colhead{Target, date} & \colhead{L1527 IRS, 26 August 2012}
}
\startdata
Coordinate Center & R.A.(J2000)=4$^\mathrm{h}$39$^\mathrm{m}$53$^\mathrm{s}$.9000 & \\
& Dec.(J2000)=26$^\mathrm{\circ}$03$^\mathrm{'}$10$^\mathrm{"}.000$\\
Frequency & 219.5603 GHz\\
Synthesized beam & 0.96$\arcsec\times0.73\arcsec(+11^{\circ}$)\\
Primary beam & 28.6$\arcsec$\\
Velocity resolution & 0.17 km s$^{-1}$\\
Noise level (detected channel) & 8.0mJy beam$^{-1}$\\
Minimum baseline & 18m \\
Maximum recoverable scale & 15.7$\arcsec$\\
Flux calibrator & Callisto\\
Gain calibrator & J0510+180\\
\hline 
\enddata
\end{deluxetable*}
\end{center}

%In this section are the photometric information, in Table \ref{tab:photometry_7.14} and Table \ref{tab:photometry_71.4}, used to fit the SEDs.

\begin{center}
\begin{deluxetable*}{cccc}
\tablenum{6}
\caption{Photometry at 7.14" aperture size.\label{tab:photometry_7.14}}
\tablehead{
%\colhead{} & \colhead{C$^{18}$O(2-1) }\\
\colhead{Wavelength} & \colhead{$F_{\lambda}$} & \colhead{Aperture$^a$} & \colhead{Reference} \\
\colhead{($\mu$m)} & \colhead{(mJy)} & \colhead{(arcsec)} & \colhead{} 
}
\startdata
$2.16$ & 0.594$\pm$0.16 & 7.14 & 1 \\
$3.6$ & 6.936$\pm$0.69 & 7.14 & 1\\
$4.5$ & 22.75$\pm$2.28 & 7.14 & 1\\
$5.8$ & 29.93$\pm$2.99 & 7.14 & 1\\
$8.0$ & 18.83$\pm$3.80 & 7.14 & 1\\
$24$ & 660.6$\pm$66 & 13 & 1\\
$70-160^b$ & 22000-64000$\pm$500-7000 & 14 & 2\\
$70^b$ & 16746.0$\pm$49.0 & 6 & 3\\
$100^b$ & 28942.0$\pm$653.0 & 6& 3\\
$160^b$  & 47011.0$\pm$16291.0 & 12 & 3\\
$1300$  & 375.0$\pm$75.0 & 6 & 4\\
$2700$  & 47 $\pm$5.6 & 3.2$^{c}$ & 5,6\\
\hline 
\enddata
\end{deluxetable*}
\end{center}

\begin{center}
\begin{deluxetable*}{cccc}
\tabletypesize{\footnotesize}
\tablenum{7}
\caption{Photometry at 71.4" aperture size.\label{tab:photometry_71.4}}
\tablehead{
%\colhead{} & \colhead{C$^{18}$O(2-1) }\\
\colhead{Wavelength} & \colhead{$F_{\lambda}$} & \colhead{Aperture$^a$} & \colhead{Reference} \\
\colhead{($\mu$m)} & \colhead{(mJy)} & \colhead{(arcsec)} & \colhead{} 
}
\startdata
%$1.66$ & 7.0$\pm$3.5 & 71.4 & TIFKAM \\
$2.16$ & 35.2$\pm$16.2 & 71.4 & 1 \\
%       & 35.2$\pm$16.2 & 71.4  & TIFKAM\\
$3.6$ & 141.8$\pm$16.2 & 71.4 & 1\\
%      & 141.84$\pm$16.2 & 71.4 & IRAC\\
$4.5$ & 225.1$\pm$16.3 & 71.4 & 1\\
%      & 225.1$\pm$16.3 & 71.4 & IRAC\\
$5.8$ & 149.5$\pm$45.0 & 71.4 & 1\\
%      & 149.5$\pm$45.0 & 71.4 & IRAC\\
%$6.7$ & 22.36$\pm$5.6 & 7.2 & ISOCAM\\
$8.0$ & 54.5$\pm$25.0 & 71.4 & 1\\
%      & 54.54$\pm$25.0 & 71.4 & IRAC\\
%$9.6$ & 1.37$\pm$0.34 & 7.2 & ISOCAM\\
%$11.3$ & 3.081$\pm$0.77 & 7.2 & ISOCAM\\
%$14.3$ & 15.3$\pm$3.8 & 7.2 & ISOCAM\\
$25$ &  743.6$\pm$70.0 & 23$\times$150$^c$ & 7\\
$60$ & 17770.0$\pm$1600.0 & 45$\times$150$^c$ & 7\\
$70-160^b$ & 22000-64000$\pm$500-7000 & 14 & 2\\
$70^b$ & 16746.0$\pm$49.0 & 6 & 3\\
$70$ & 24170.0$\pm$4834.0 & 75 & 1\\
$100$ & 73260.0$\pm$11700.0 & 90$\times$150$^c$ & 7\\
$100^b$ & 28942.0$\pm$653.0 & 6 & 3\\
$160^b$  & 47011.0$\pm$16291.0 & 12 & 3\\
$160$  &  94000.0$\pm$38000.0  & 60 & 8\\
$350$  & 44000.0$^{d}$$\pm$20000.0$^{d}$ & 45-60$^{e}$ & 1\\
$450$  & 33125.0$^{d}$$\pm$20900.0$^{d}$ & 40-120$^{e}$ & 1\\
$750$  & 8400.0$\pm$1100.0 & 45 & 9\\
$800$  & 1400.0$\pm$560.0 & 60 & 8\\
$850$  & 6167.0$^{d}$$\pm$480.0$^{d}$ & 40-120$^{e}$ & 1\\
$1300$  & 1110.0$^{d}$$\pm$110.0$^{d}$ & 30-40$^{e}$ & 1\\
$2700$  & 47.0$\pm$5.6 & 3.2$^c$ & 5,6\\
\hline 
\enddata
%\multicolumn{4}{@{}l@{}}{\scriptsize *$X$ is the fractional abundance relative to hydrogen, $n_{X}/(n_{H}+2n_{H{_2}})$.} \\
\tablecomments{$^{a}$radius. $^{b}$The apertures of the \textit{Herschel} HPPSC catalog and CDF archive spectrum generally lie} between the 7.14" and 71.4" aperture values, thus, the same \textit{Herschel} data are shown on both SED plots. $^c$radius of the cited beam; either the Full Width Half Maximum (FWHM) size or geometric mean, divided by two. 
$^{d}$For a given wavelength $>$200$\micron$, the average of values that are listed in \citep{Tobin_2008} and largest error fluxes are adopted. 
$^{e}$ range of contributing apertures. 
References:
$^1$\citep{Tobin_2008},
$^2$\citep{Green_2016},
$^3$\citep{Herschel_2017},
$^4$\citep{Motte_2001},
%$^5$\citep{Yen_2013},
$^5$\citep{Ohashi_1997},
$^6$\citep{Terebey_1993}
$^7$\citep{Beichman_1988},
$^{8}$\citep{Ladd_1991},
$^{9}$\citep{Chandler_2000}
\end{deluxetable*}

%\begin{tablenotes}
%\item 

%\end{tablenotes}
\end{center}

%References

\bibliography{L1527-1}

\end{document}